# Improving Today, Narrowing Tomorrow: Collective Learning, Diversity, and Generativity

Esteve Almirall esteve.almirall@esade.edu Christopher Tucci c.tucci@imperial.ac.uk

## Abstract

Generative AI makes a paradox newly visible: learning from a common source can improve what each firm does today while narrowing the variety available for tomorrow's discoveries. The tension extends beyond AI. Fields turn recurring features of successful organizations into portable practices—just-in-time production, freemium business models, or mixture-of-experts architectures. These collective abstractions are partial templates others can adapt, not designs or universal prescriptions. We study how they shape discovery in a model of firms searching interdependent landscapes. A collective repertoire extracts practices from patterns shared by leading configurations; firms vary in how they evaluate those practices locally. The model reveals a generativity chain: population diversity supplies heterogeneous experience, collective abstraction turns it into reusable options, and situated judgment matches options to local needs. Collective learning is not inherently homogenizing. Copying leaders consumes diversity quickly, whereas repertoires paired with local judgment preserve generative capacity. Environmental disruption exposes the consequences: it devalues accumulated knowledge but reopens local search, helping homogeneous, exhausted industries while harming diverse industries whose repertoire remains useful—even though diverse industries still perform better overall. Recurring practices may also outlive their usefulness, making popularity a poor test of validity after change. These findings matter for organizations using generative AI, best-practice communities, and research fields converging on dominant strands. Generativity requires governing the diversity feeding knowledge and the judgment applied to it. Learning systems should be judged not only by the answers they spread today, but by whether they preserve the differences from which tomorrow's answers can be made.

## 1. Introduction

Generative in name, convergent in use? Generative AI can improve what each firm produces while narrowing what a field produces collectively. In an experiment, writers given access to AI-generated ideas produced stories judged more creative, yet their stories became more similar to one another (Doshi and Hauser 2024). Concentration is also visible in the technology's supply. The UK Competition and Markets Authority mapped more than 90 partnerships and strategic investments involving the same six foundation-model firms and warned that dependence on a small set of providers may reduce choice and innovation (Competition and Markets Authority 2024). The managerial tension is immediate: drawing on a common source may help each organization today while depleting the variation from which tomorrow's best practices will be collectively abstracted.

This contemporary concern reveals a broader puzzle: can the same disruption raise frontier discovery in one industry while reducing it in another? A common intuition is that industry heterogeneity buffers environmental disruption by spreading exposure across distinct configurations. If firms operate differently, a change that undermines one configuration may leave others viable; if firms have converged, the same change appears likely to damage them together. This portfolio logic informs how managers discuss technological discontinuities and how policy makers evaluate concentrated operating models. It also fits a broader view of diversity as insurance against an uncertain environment.

The model developed here identifies an important boundary condition to that intuition. Holding the environment, search process, resources, and response constant, the pooled effect of an identical disruption on terminal frontier discovery is positive in a low-diversity industry and negative in a high-diversity one. The difference is not preparedness after the event. It is the adaptive state that prior learning created before the event. Diversity is both the asset and the exposure: it supplies the varied experience from which a useful repertoire is collectively abstracted, and that accumulated usefulness is precisely what a disruption can depreciate. Low-diversity industries have less knowledge at risk, but they are also more likely to have exhausted local improvements.

This sign reversal concerns the effect of the disruption, not the level of performance. Diversity remains advantageous in every condition we study. A complete redrawing of payoffs removes roughly half of the performance advantage associated with diversity, but it does not reverse the ordering: the diverse industry exposed to a complete disruption still finishes ahead of the homogeneous industry that experiences no disruption. Understanding that distinction is central to the argument.

We explain the pattern through collective generativity: an industry's capacity to keep discovering improvements from differences already present among its firms and in its shared repertoire. In the model, best practices are partial templates built from decision patterns that recur among several leading configurations. They represent ideas such as just-in-time production, freemium business models, or mixture-of-experts architectures - not complete firm designs or universal prescriptions.

The argument follows three kinds of variation. Population diversity is how much firms' configurations differ. Source diversity is how much the leading configurations differ. Option diversity is how differently available practices would perform for a particular firm. Situated judgment means comparing those local consequences before adopting. We call the best configuration found by any firm discovery. These distinctions let us trace how variation becomes shared knowledge and then further discovery.

Disruption affects this system through two opposing processes. The depreciation process reduces the current value of collective knowledge: practices abstracted before the disruption become less predictive of local fit. The reopening process changes the search landscape itself: firms that had reached local peaks may again find improving moves. Reopening occurs across diversity conditions, but its conversion into frontier discovery depends on interdependence, available search resources, and the content firms can evaluate. Pre-disruption diversity therefore shapes both what is exposed to depreciation and the conditions under which renewed search creates value.

The depreciation process also reveals an information problem. The collective abstraction process recognizes best practices because their elements recur among leading distinct configurations, not because any single actor continuously tests their causal value. A disruption can therefore reduce a practice's usefulness without immediately reducing its prevalence. Even after a complete disruption, fewer than one in twelve published practices loses support, while the option diversity of the repertoire falls by more than 40%. When a field audits knowledge by continued adoption, a stale practice can still look valid.

Together, these processes yield the paper's central claim: whether destruction is locally creative depends on the adaptive state produced by prior learning. A disruption can reopen search in a field that has depleted its generative variety, but it can destroy more repertoire value than renewed search recovers in a field whose collective knowledge remains useful. The relevant question is not simply how large the disruption is, but what the industry has already learned itself into.

## The setting

We study 100 firms searching performance landscapes defined by 48 interdependent yes-or-no decisions, five levels of interdependence, and 400 rounds. Each firm first tries to improve by changing one decision.

Only when no one-decision change helps may it learn socially. All firms receive the same amount of time and the same allowance for imitation.

The four learning architectures vary two features. First, a candidate comes either from one observed peer or from a pattern recurring across several leading configurations. Second, the adopter either applies that candidate directly or first evaluates its local fit (Rivkin 2000; Csaszar and Siggelkow 2010; Posen, Lee and Yi 2013). Our focal arrangement combines a public repertoire of partial best-practice templates with evaluation by the adopting firm. We first establish this arrangement's value, then show where that value comes from, how imitation changes its inputs, and how disruption alters it.

The modeled disruption is an exogenous environmental disruption: at round 200, the contribution tables of a randomly selected fraction φ of loci are redrawn and, in the reported variant, their dependencies are rewired. This operation captures rapid depreciation of prior knowledge, not the full set of features associated with technological disruption such as entry, demand migration, substitution, or incumbent exit. The correlation between pre- and post-disruption landscapes is 1 - φ, an analytic prediction confirmed in the implementation. At φ = 1, the post-disruption payoff landscape is independent of the one on which firms previously learned.

### What we find

The first results explain why comparing alternatives matters. Firms gain when available practices fit their own configurations differently; simply screening practices by how well their source firms performed does not create the same value. This result is similar across three continuously updated catalogues built from top performers, performance-screened bundles, or random sources. In an early-frozen boundary condition, the gain nearly disappears because the old source configurations soon cease to outperform adopters, leaving almost no eligible bundles.

Those useful differences among options ultimately come from population diversity. At N = 48, reducing population diversity to its lowest modeled level lowers discovery by about thirteen times as much as replacing top-performer sources with random sources; the ordering is the same at N = 32 and N = 64. Situated judgment is valuable because a diverse population supplies a repertoire of candidates that fit firms differently.

The chain is also endogenous. The value of judgment remains private to firms that possess it, and substituting a industry-wide mandate for local evaluation reduces performance. At the same time, adoption changes the population from which future practices are abstracted. Copying the best observed peer reduces population diversity by 0.140 over the second stage of the model, whereas locally selecting from a collective catalogue reduces it by 0.060 under the same horizon and social-learning allocation. Complete learning architectures therefore differ in how quickly they consume the variation that supports future learning.

The second results show what disruption changes. A complete landscape redraw removes 98-100% of firms from their old local peaks, creating nearby improvement opportunities. At the same time, it changes the value of catalogue content. Relative to matched random movement, locally selected catalogue content becomes more valuable after disruption in low-diversity industries and less valuable in high-diversity industries.

Across all four ways of limiting search, the average effect on frontier discovery shifts from positive to negative as prior diversity rises. This result has an important boundary: under the tightest limit, disruption does not improve the low-diversity frontier at two interdependence levels. Diversity nevertheless produces the higher final frontier in every tested disruption condition. The claim is therefore conditional, not a claim that displacement is generally productive or that low diversity is desirable.

### Contribution

The paper's principal contribution is a conditional account of creative destruction grounded in an endogenous adaptive state. Prior learning determines both what an environmental disruption destroys and what it makes newly reachable. The same population diversity that builds a useful collective repertoire also creates exposure when that repertoire is depreciated; the same convergence that weakens generativity can create scope for renewed movement when disruption occurs. This explains why an identical disruption can have opposite effects without reversing the performance advantage of diversity.

A supporting contribution specifies the multilevel mechanism that produces this state. Population diversity supplies heterogeneous configurations; collective abstraction converts recurring elements among leading distinct configurations into portable practices; option diversity gives situated judgment something consequential to select; and imitation changes the upstream population. This mechanism links research on organizational search, knowledge transfer, and collective learning while sharply bounding the monitoring implication: prevalence can lag usefulness when repertoires are audited through recurrence rather than current performance tests.

Figure 1 maps the motivating AI example onto the modeled learning system and separates the two disruption processes from the performance-level result.

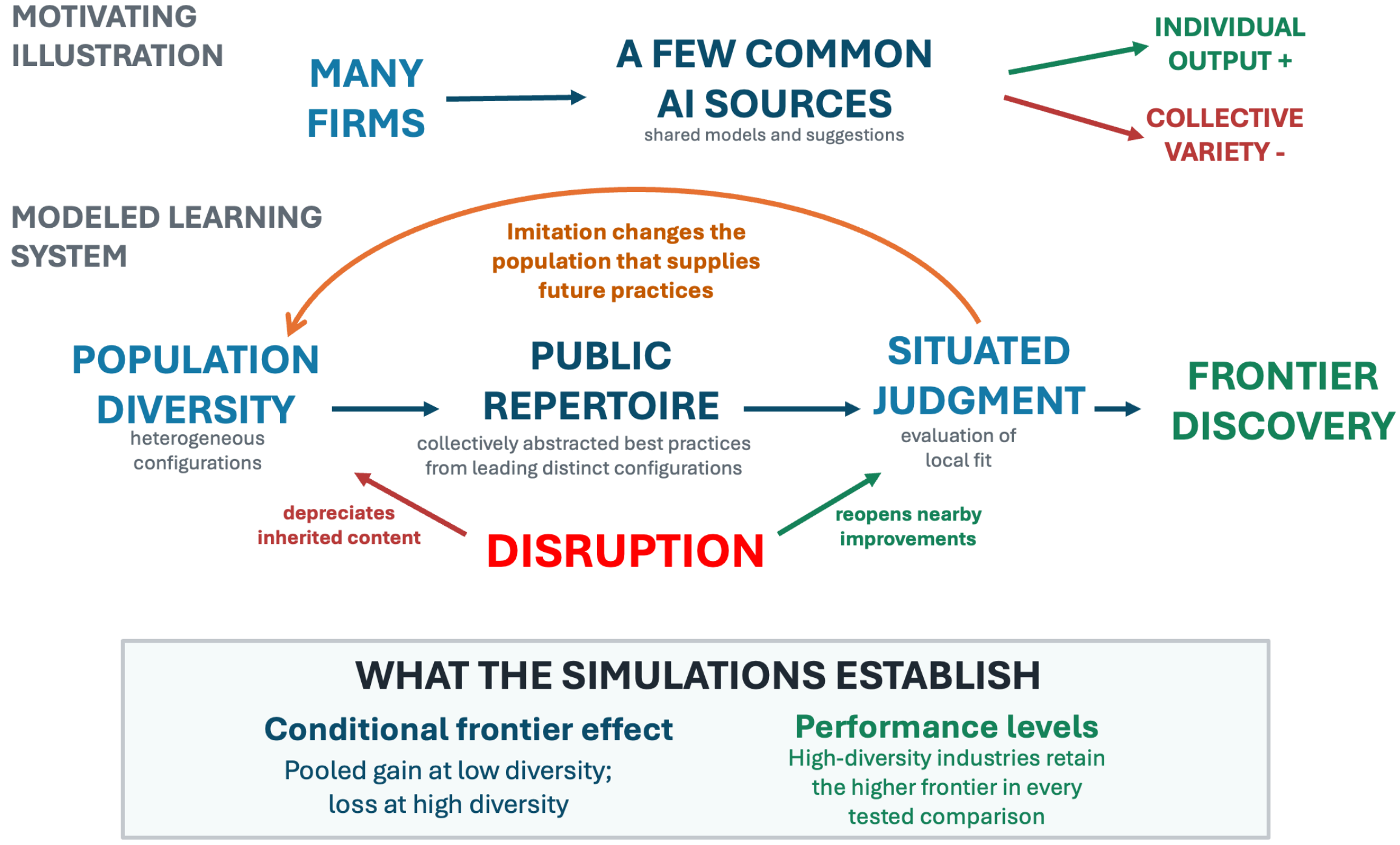

*Figure 1. From generative AI to collective generativity. Common AI sources illustrate how individual gains can coexist with lower collective variety (Doshi and Hauser 2024) amid concerns about foundation-model concentration (Competition and Markets Authority 2024). The model generalizes this tension: population diversity feeds collectively abstracted best practices; situated judgment converts optionality into discovery; and disruption depreciates inherited content while reopening nearby improvements. The pooled low-diversity frontier gain weakens under the tightest hard limit on changed decisions; diversity retains its performance-level advantage throughout.*

## 2. Theory

### 2.1 Population diversity and option diversity

Diversity in problem solving is usually modeled as a property of searchers. Hong and Page (2004) show that heterogeneous problem solvers can outperform a homogeneous group of higher-ability solvers because they bring different local moves and therefore become trapped in fewer common locations. In this account, diversity broadens the search conducted by a group and can substitute, under specified conditions, for individual ability.

We distinguish population diversity from option diversity. Population diversity describes how firms in an industry differ in their operating configurations. Option diversity describes how candidate practices differ in the value they would create when combined with a particular adopter's configuration. The two constructs occupy different positions in the causal chain. Population diversity is a upstream input to collective knowledge; option diversity is a property of what an individual firm can evaluate.

This distinction changes the relation between diversity and ability. Diversity among searchers can substitute for individual ability by expanding the moves attempted by the group. Diversity among options instead complements evaluation ability. If all candidate practices have the same local fit, evaluating more candidates cannot improve selection. The value of judgment requires consequential variation among the alternatives.

The logic is straightforward. With one candidate, a firm has nothing to compare. With m candidates, it chooses the one that performs best in its own configuration. Raising every candidate's payoff by the same amount raises the outcome but does not change the gain from comparing them. Widening the differences among their local payoffs does. Appendix C states this argument formally as an order-statistic result.

A performance screen can also change the spread of local payoffs, so the theory does not claim that screening is irrelevant. It makes a narrower prediction: given a usable set of alternatives, raising all candidates' average payoff need not raise the return to comparing them, whereas widening their differences in local fit should. A screen based on performance elsewhere may even remove some of the variation that makes situated evaluation useful.

The first link in the argument is therefore a complementarity. Situated judgment creates more value when the collective repertoire offers options with heterogeneous local fit. The next question is where that heterogeneity originates.

### 2.2 Collective generativity and the architecture of imitation

In the model, option diversity is not an exogenous feature of the repertoire. Best practices are collectively abstracted from recurring elements of leading distinct configurations, so the repertoire can contain only differences represented in the field. Population diversity is thus the raw material from which option diversity is produced. A homogeneous population may support widely used practices, but it cannot supply many substantively different candidates for local evaluation.

This account extends organizational-learning research that separates access to knowledge from the capacity to use it (Argote, Lee and Park 2021; Lee and Van den Steen 2010). Transfer templates can make knowledge portable, but their value still depends on contextual fit and adaptation by the recipient (Winter and Szulanski 2001; Jensen and Szulanski 2007). Our added step is to make the variety inside the transferable repertoire endogenous to the configurations represented in the field. This distinction is also central to crowdsourcing: broadcasting a problem can transform distant search into local search, while leaving the seeker with the task of evaluating heterogeneous solutions (Afuah and Tucci 2012).

This relation places diversity among searchers upstream of diversity among options. Different firms supply different configurations; the abstraction process turns some recurring differences into portable practices; situated judgment selects among those practices; and the selected practice may enable further discovery. Not every population difference survives abstraction, because only recurring patterns among leading distinct configurations enter the catalogue. The variation present in the population therefore limits the variation available to firms later.

This constraint matters for collective knowledge production. Under recurrence-based abstraction, improving the average quality of the repertoire cannot fully compensate for a population that supplies little variation. Entrepreneurs, managers, consultants, professional communities, and scholars can identify, name, compare, and update practices; collectively, however, they cannot recover heterogeneous operating experience that is absent from the field. Research on competing innovation communities likewise shows that free revealing and knowledge brokering can improve average system learning while creating a trade-off with the best attainable outcome (Villarroel, Taylor and Tucci 2013).

The same process can alter its own input. March (1991) showed that learning between individuals and an organizational code can reduce the heterogeneity from which the code learns. Here the repertoire is mined endogenously rather than stipulated. When firms adopt practices extracted from leading configurations, their own configurations move toward the population that supplied those practices. Subsequent catalogs are then abstracted from a less varied field.

How quickly diversity declines should depend on the complete learning architecture. Copying the best visible peer directs adopters toward a common target; drawing from a collectively abstracted repertoire and evaluating local fit can send adopters toward different practices. Posen, Lee and Yi (2013) show how imperfect copying can preserve heterogeneity. Our model shifts attention to the joint arrangement of sourcing, bundling, and evaluation. Because the reported comparison changes these dimensions together, it establishes different rates of depletion across architectures but does not attribute the difference to collective sourcing alone.

We call the resulting capacity collective generativity: an industry's ability to continue discovering improvements from variation already present in its firms and shared repertoire. Population diversity supplies the initial variation. Diversity among the leading source configurations determines what can be abstracted. Differences in how candidate practices fit a particular firm give situated judgment something to select. Available improving moves reveal remaining search opportunities, and discovery records the best configuration the industry reaches. Generativity differs from resilience, which concerns maintaining or recovering performance after adversity, and from exploration, which names an activity rather than the remaining capacity for useful variation. It also differs from platform generativity, which emphasizes an ecosystem's capacity to produce complements under governance tensions (Cennamo and Santaló 2019).

The shield-engine distinction follows from this definition. A shield would reduce the damage produced by a disruption. A generative resource instead allows a field to keep improving before any disruption occurs. Diversity may therefore increase the amount of collective knowledge exposed to depreciation while still leaving the diverse industry better off in levels.

Generative artificial intelligence illustrates, but does not test, the distinction. Doshi and Hauser (2024) find that access to a common generative source improves individual creative output while increasing similarity across outputs. In the language of the model, concentration on a common source resembles exemplar-based imitation; a repertoire extended by many contributors and evaluated locally resembles collective sourcing. The implication concerns how a field organizes access and evaluation, not an empirical claim about the effects of any specific AI system.

### 2.3 Situated judgment as a private capability

The second link in the chain is the adopting firm's evaluation. Knudsen and Levinthal (2007) distinguish generating alternatives from evaluating them. The catalogue performs the first function by supplying candidate practices; the firm performs the second by assessing each candidate in its own configuration. Situated judgment is the capacity to make that local comparison before committing to adoption.

Because the relevant payoff is a match between practice and configuration, the capability should be private in two senses. Its return need not spill over to firms that lack evaluation ability, and a community-level choice cannot substitute for local assessment. A practice that fits one configuration may not fit another even when every firm observes the same public catalogue.

Increasing the share of capable firms can therefore raise the industry mean by changing the composition of firms that select well, without improving the outcomes of incapable firms. This prediction differs from a knowledge-access account, under which capable firms would improve a public stock that all firms could use. Here access is public; evaluation remains firm specific.

For the same reason, forcing all firms to adopt the most widely supported practice should reduce performance when local fit varies. The cost is not necessarily homogenization. Firms can respond to a common misfit by repairing in different directions, thereby becoming more heterogeneous even as performance falls. The theoretical loss is the removal of matching discretion, not an inevitable reduction in population diversity.

## 2.4 Disruption as depreciation and reopening

Models of search under turbulence emphasize that disruption depreciates accumulated knowledge and alters the exploration-exploitation balance (Posen and Levinthal 2012; Siggelkow and Rivkin 2005). We retain that depreciation process but add a countervailing possibility. When prior search has exhausted locally improving moves, a disruption can recreate an improvement gradient. Disruption may therefore both remove useful knowledge and expand the set of reachable gains.

The depreciation process operates through the collective repertoire. Practices enter the catalogue because they recur among leading distinct configurations before the disruption. Redrawing payoffs can eliminate their ability to predict local fit without changing their immediate prevalence. A recurrence-based catalogue can remain popular while becoming less useful. The return to situated judgment should decline with the magnitude of disruption because the evaluated options encode a weaker signal.

This mechanism creates a monitoring problem with a specific boundary. It applies to repertoires audited through continued use or support rather than through direct, current tests of performance. Denrell (2003) shows how learning from observed survivors can misidentify the causes of success. Our mechanism concerns the record of practices rather than firms: prevalence survives because the disruption changes payoffs before it changes observed configurations.

The reopening process operates through local search. A firm at a local peak has no improving one-step move. Redrawing contributions can make neighboring configurations attractive again. The mechanism does not require the disruption to disperse firms across the landscape; it requires the disruption to alter the payoff gradient around their current positions.

The logic is related to Almirall and Casadesus-Masanell's (2010) result that partial adoption can displace a searcher from a local peak and renew discovery. We move that mechanism to the industry level: an environmental disruption can dislodge many firms simultaneously by changing the value of their neighboring configurations. The model therefore asks when renewed reachability compensates for the collective knowledge that the same event depreciates.

Two accounts must therefore be distinguished. A scattering account predicts that disruptions help by increasing variation among firms and that greater post-disruption dispersion should accompany greater discovery. A reopened-opportunity account predicts that disruptions help by making improving neighbors available again and that greater renewed movement should accompany greater discovery. Section 5 compares these observable implications while recognizing that descriptive associations do not prove mediation.

The relative strength of the two processes should depend on pre-disruption diversity. A diverse industry supports a repertoire with varied local fit, giving depreciation more value to remove. A homogeneous industry supports fewer consequential alternatives and is more likely to have exhausted similar local moves, giving reopening more scope.

The net effect of a disruption should consequently decline as pre-disruption diversity rises. At low diversity, the value created from reopened local opportunities can exceed repertoire depreciation; at high diversity, repertoire depreciation can dominate. The model thus predicts a crossing rather than a uniformly positive or negative effect of disruption.

The two accounts also imply different managerial levers. If scattering generated the gain, a field might reproduce it by mandating experimentation or otherwise dispersing firms. If reopened local opportunity generated the gain, variation alone would not suffice; the missing object would be a set of reachable improvements in the environment. The model distinguishes these interpretations but does not make disruption itself a controllable strategic instrument.

This conditional logic links prior learning to the consequences of turbulence. The same process that improves firms before a disruption can alter both the value exposed to depreciation and the improvements that remain reachable. The effect of disruption is therefore partly endogenous to the industry's learning history even though the disruption is exogenous.

The argument also complements work showing that formal structure guides how organizational networks regenerate (Clement and Puranam 2018). Here the pre-event learning architecture leaves behind an adaptive structure—a distribution of configurations and a public repertoire—that governs what can regenerate after disruption.

We express the condition in realized population diversity rather than in the initialization parameter. The parameter is a model device whose mapping to realized configurations varies with problem size. Diversity is the theoretically relevant state variable and the quantity that could, in principle, be compared across settings.

### 2.5 Conditional creative destruction

Creative destruction combines loss and renewal. Our argument specifies why those components need not have the same balance in every industry. The destructive component removes value embedded in a collective repertoire; the creative component reopens local improvement opportunities. Pre-disruption diversity determines how much each component can contribute under the modeled search conditions.

A disruption is more likely to raise terminal frontier discovery when prior learning has left firms similar and locally exhausted. It is more likely to reduce frontier discovery when firms remain varied and continue to draw value from their collective repertoire. This is a statement about the pooled change caused by the disruption, not an endorsement of homogeneity.

The level comparison remains decisive. If diversity sustains generativity before the disruption, a diverse industry can lose more from disruption and still finish ahead. A pooled positive frontier effect at low diversity indicates recovery from a weaker adaptive state, not superiority over the diverse industry.

The theoretical proposition is therefore conditional: disruption becomes locally creative when the value created from reopened local opportunities exceeds depreciation of the collective repertoire. Diversity shifts that balance toward depreciation while continuing to raise performance levels through the generative process that preceded the disruption.

## 3. Model

This section describes the model primitives, their mapping to the theoretical constructs, the experimental manipulations, and the comparisons used in the results. Appendix A provides the formal specification; Appendix B defines the abstraction operator and its bound; Appendix C states the order-statistic result; Appendix D defines the disruption; and Appendices E-F report estimation details, robustness analyses, and registered predictions.

### 3.1 Firms, landscapes, and search

An industry contains M = 100 firms. Each firm is represented by N yes-or-no decisions; the complete set of its decisions is its configuration. Performance is defined on an NK landscape (Kauffman 1993; Levinthal 1997). Each decision position, or locus, contributes a value that depends on its own setting and on K other decisions. Firm performance is the average of those contributions. Larger K means that more decisions interact, making the landscape more rugged and local improvements harder to combine. All firms in a run face the same landscape but begin from different configurations.

The main analysis uses N = 48 and five values of K spanning nearly separable to nearly fully coupled landscapes. We replicate the central results at N = 32 and N = 64. N denotes the number of loci; M denotes the number of firms and remains 100.

Figure 2a summarizes a round. A firm first tries a one-decision change and keeps it if performance improves. If no neighboring configuration performs better, the firm is at a local peak and may consult a social source. The simulation has two 200-round stages. In each stage, imitation has a nominal allowance equivalent to changing 20 decisions; adopting a multi-decision bundle uses more of that allowance than adopting a short bundle. In the archived benchmark, this allowance can block a social adoption but does not interrupt an improving one-decision move, so the realized total can slightly exceed 40. Section 5.6 tests hard limits that apply to every change. All conditions otherwise have the same simulation length and applicable imitation allowance. Discovery is the highest-performing configuration found by any firm.

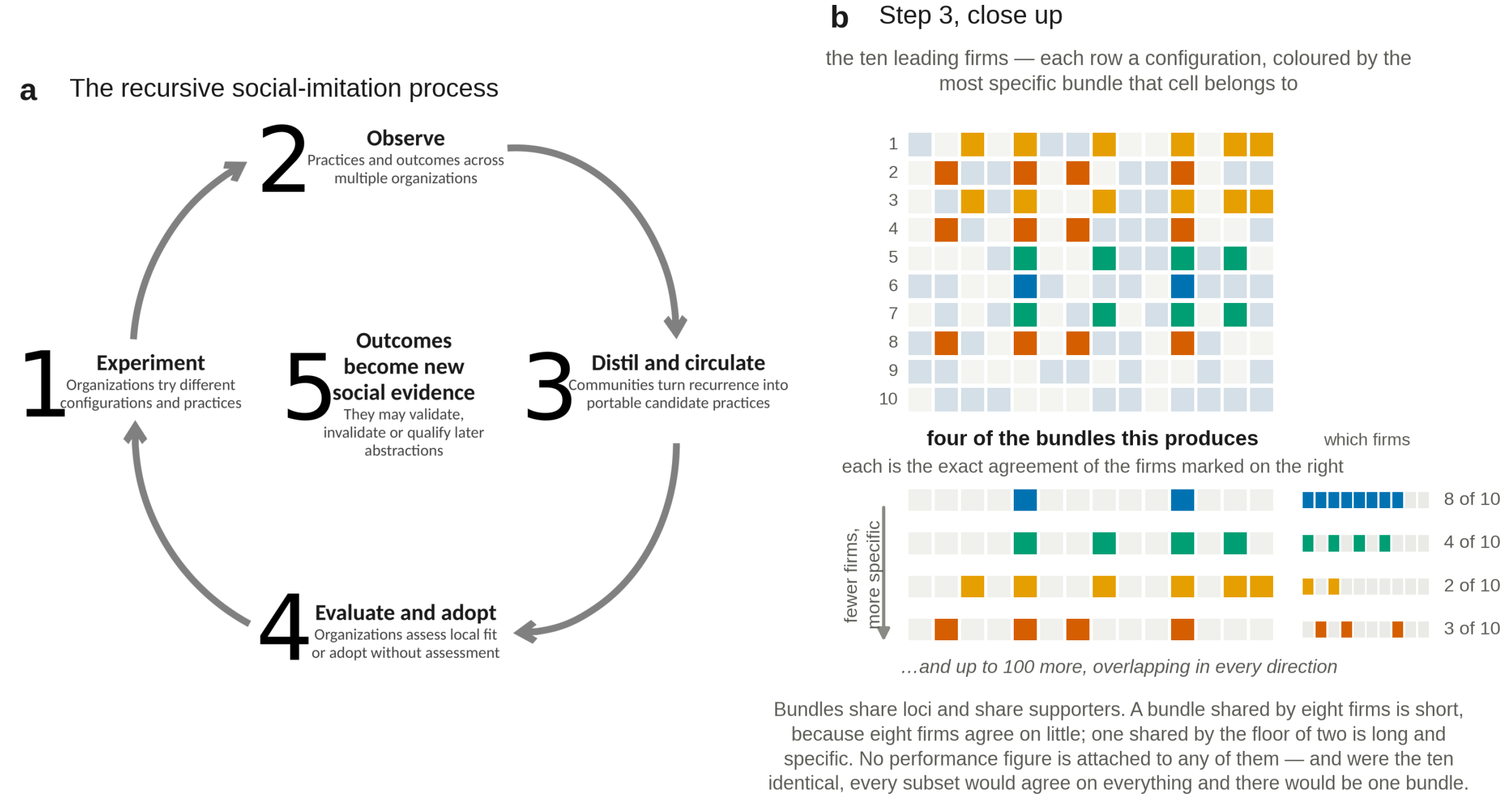

*Figure 2. The modeled learning process. (a) Firms alternate between one-decision local improvement and social learning under the same simulation length and allowance for imitation. (b) The collective abstraction rule identifies overlapping bundles, each consisting of the decisions on which a particular group of leading distinct configurations agrees. Configurations, not firm identities, are the source unit. Bundles vary in length and in the number of sources that support them; up to 100 are published at once, and a bundle carries no estimate of how well it will perform for an adopter.*

### 3.2 Four architectures of social learning

The four social-learning conditions cross the source of a candidate with the adopter's decision rule. A candidate comes either from one observed peer or from a pattern recurring across several firms. The adopter either applies the candidate without comparing its local consequences or evaluates its fit before adoption. After a social adoption, the firm resumes one-decision local improvement; we call these follow-on adjustments accommodation or repair.

Copy-the-best is the dyadic, blind benchmark. A firm observes the configurations and realized performance of ten fixed network neighbors and copies part of the best neighbor's configuration (Rivkin 2000; Posen,

Lee and Yi 2013). The adopter observes total performance but cannot identify which copied choices produced it.

Selective copying retains the dyadic source but adds evaluation: the firm compares several subsets of one peer's configuration in its own setting and adopts the best improving candidate (Csaszar and Siggelkow 2010).

Blind social imitation adopts a practice drawn from the socially abstracted catalogue without evaluating local fit.

Informed social imitation, the focal architecture, combines the collective catalogue with situated judgment. A firm draws up to m eligible catalogue bundles, calculates the performance each would produce after local repair, and adopts the best candidate it can afford. The parameter m is the maximum number of candidates compared at once. When $m = 1$, the firm sees one candidate and therefore cannot choose among alternatives.

These four conditions map the source and decision-rule dimensions within the same model. To isolate the value of situated selection, the focal comparisons match how often firms adopt and how many decisions they change. Broader comparisons show how complete learning arrangements deplete diversity at different rates. Because some arrangements change the source, bundle construction, and evaluation rule together, they do not identify the effect of sourcing alone.

### 3.3 Collective abstraction of best practices

The model represents the collective process through which a field turns recurring elements of leading distinct configurations into portable best practices. Each modeled practice is a bundle that specifies values for only some of the firm's decisions, not a complete design. In what follows, catalogue denotes the model's current list of published bundles; collective repertoire denotes the broader stock of shared practices that this list represents. In empirical settings, entrepreneurs, managers, consultants, professional associations, scholars, and standards communities may all participate in abstraction. Familiar examples include just-in-time production, freemium business models, and mixture-of-experts architectures. The model compresses this distributed process into a transparent rule; it does not posit a single knowledge producer.

At each round, the abstraction rule ranks all configurations in the population by performance, removes duplicates, and selects up to $E = 10$ leading distinct configurations. For every group of at least two selected configurations, it records the largest bundle of decision values on which all members agree. Duplicate bundles are retained only once. A bundle's support is the number of selected source configurations that contain it. Figure 2b illustrates this computational representation of collective abstraction; Appendix B gives the formal rule and a worked example.

The catalogue is therefore not a single consensus practice. It contains overlapping bundles supported by different groups of source configurations, with up to 100 current entries. Because different groups agree on different decisions, firms using the same public catalogue may evaluate different candidates. The bundle shared by every source is only one, typically short, entry among many.

Three properties connect this catalogue construction rule to the theory.

First, the abstraction rule records practice elements, not performance estimates. A firm learns a bundle's value only by evaluating the configuration that adoption would produce. This feature separates public access from situated judgment.

Second, the catalogue can publish only combinations already represented among the leading source configurations. With E sources, the number of distinct bundles is bounded by the number of nonempty source groups: at most $2^E - 1$, regardless of how many decisions each configuration contains (Appendix B). When the leading configurations resemble one another, many source groups yield the same bundle, so the catalogue offers few meaningfully different options.

Third, publication depends on recurrence, not a current performance test. A bundle remains published while enough current leaders exhibit it; the catalogue construction rule does not test whether adopting that bundle would still improve a firm. The popularity-usefulness distinction in Section 5 follows from this information rule and should not be generalized to repertoires that directly test current performance.

### 3.4 Manipulations and measures

The diversity of the population. To create industries that begin with different degrees of similarity, the model starts from one reference configuration and varies r, the maximum number of decisions on which a firm's initial configuration may differ from that reference. This number of differing decisions is the Hamming distance. Small r produces near-identical firms; the standard draw produces essentially unrelated configurations. We report realized population diversity - the share of firms occupying distinct configurations before disruption - rather than r. The same r creates different amounts of realized diversity at different problem sizes: r = 2 yields shares of 0.22, 0.40, and 0.48 at N = 32, 48, and 64.

Catalogue construction varies which source configurations and which mined bundles enter the published list. Top-performer source selects the ten highest-performing distinct configurations and publishes recurring bundles that meet the support rule. Performance-screened begins with those same candidate bundles, then compares the average performance of source configurations containing each bundle with the average performance of sources without it. Each group must contain at least three configurations, and the bundle is published only when the first average is at least as high as the second. This is an association screen, not evidence that the bundle caused source performance or will fit a particular adopter. Random source applies the same recurrence rule to ten randomly selected distinct configurations.

All catalogue conditions then use the same adopter-specific eligibility rule. A published bundle can be considered only if it would change at least two of the adopter's decisions and if the configurations that supplied it performed better on average than the adopter currently does. The early-frozen benchmark fixes the catalogue once it fills during the initial rounds, before firms have learned much. The list still exists, but as firms improve its early sources rarely remain better than current adopters, so almost no bundles remain eligible. This is a boundary check with an effectively empty usable choice set, not a fourth substantive curation design.

Evaluation capacity, m, is the maximum number of eligible catalogue bundles a firm can compare at one decision point. It takes six values: $m = 1, 2, 3, 5, 8$, and 10. Separate analyses vary the share of firms able to compare ten candidates.

The environmental disruption occurs at round 200. A fraction $\varphi$ of loci receives newly drawn contribution tables; in the reported variant, the corresponding dependency sets are also rewired. The operation captures abrupt depreciation of knowledge about payoffs and architecture. It does not model entry, exit, strategic response, demand change, or incumbent displacement. The pre- and post-disruption landscape correlation is $1 - \varphi$, derived analytically and confirmed in the implementation (Appendix D). Results are unchanged when payoffs are redrawn without rewiring (Appendix F).

Catalogue inertia, $\lambda$ (lambda), is the fraction of the pre-disruption catalogue deliberately carried into the next round with its old recorded support. The main disruption analyses set $\lambda = 0$, so the catalogue is rebuilt from current configurations and the effect of disruption is not confounded with deliberate retention. Section 5.2 varies inertia to test what happens when old practices remain published.

### 3.5 Comparators, outcome criteria, and statistics

The analyses use three safeguards to separate mechanisms and avoid mean-only interpretations.

First, we separate useful catalogue content from the effect of simply moving a firm away from its current local peak. For every locally evaluated catalogue condition, a matched random-movement condition changes decisions at the same realized rate and in bundles of the same size, but chooses the decision positions and values randomly rather than from the catalogue. Their difference in average firm performance is the catalogue-content advantage: what selected repertoire content adds beyond comparable displacement. We examine the value of evaluation capacity separately by comparing the best solution discovered when firms can evaluate ten catalogue options with the best solution discovered when they see only one.

Second, the primary outcome is discovery - the best configuration found by any firm. Average firm performance and population diversity are separate secondary outcomes. An unmatched random-adoption condition can raise the average while reducing the best solution found and collapsing diversity, so we do not call a mean-only improvement a discovery gain. Keeping the outcomes separate prevents distinct theoretical quantities from being folded into one index.

Third, every comparison pairs conditions that use the same landscape, starting population, network, and disruption draw. These common inputs are generated from the same seed, problem size N, interdependence K, and replication number. We report 95% Student-t intervals with n - 1 degrees of freedom. We verify matched movement before comparing outcomes, and Appendix F reproduces the predictions that were registered before the simulations were interpreted.

## 4. Results I - the generativity chain

Unless stated otherwise, this section reports static-landscape results for N = 48. Contrasts are paired within landscape and replication; intervals are 95% Student's t intervals with n - 1 degrees of freedom.

### 4.1 Judgment creates value by choosing among options that fit differently

Does judgment create value because a catalogue contains practices associated with stronger source performance, or because it gives firms alternatives that fit their situations differently? Figure 3 compares three continuously updated catalogue designs. Top-performer source mines recurring bundles from the ten leading distinct configurations. Performance-screened starts with the same bundles but publishes one only when the source configurations containing it perform at least as well, on average, as those without it; at least three configurations are required in each group. For example, a bundle appearing in four leaders is retained only if those four leaders perform at least as well on average as the six leaders without it. Random source mines recurring bundles from ten randomly selected distinct configurations. The early-frozen boundary condition fixes the catalogue during the initial rounds rather than updating it.

The catalogue determines what a firm may consider, not what it must adopt. A published bundle is eligible for a firm only when it would change at least two of that firm's decisions and its source configurations performed better on average than the firm currently does. When the firm becomes stuck in local search, it draws up to m eligible bundles, evaluates the performance each would produce in its own configuration after repair, and selects the locally best candidate. Figure 3a reports the average-performance advantage over matched random movement at m = 1, 2, 3, 5, 8, and 10. Figure 3b reports a different outcome: the increase in the best solution discovered when evaluation capacity rises from m = 1 to m = 10.

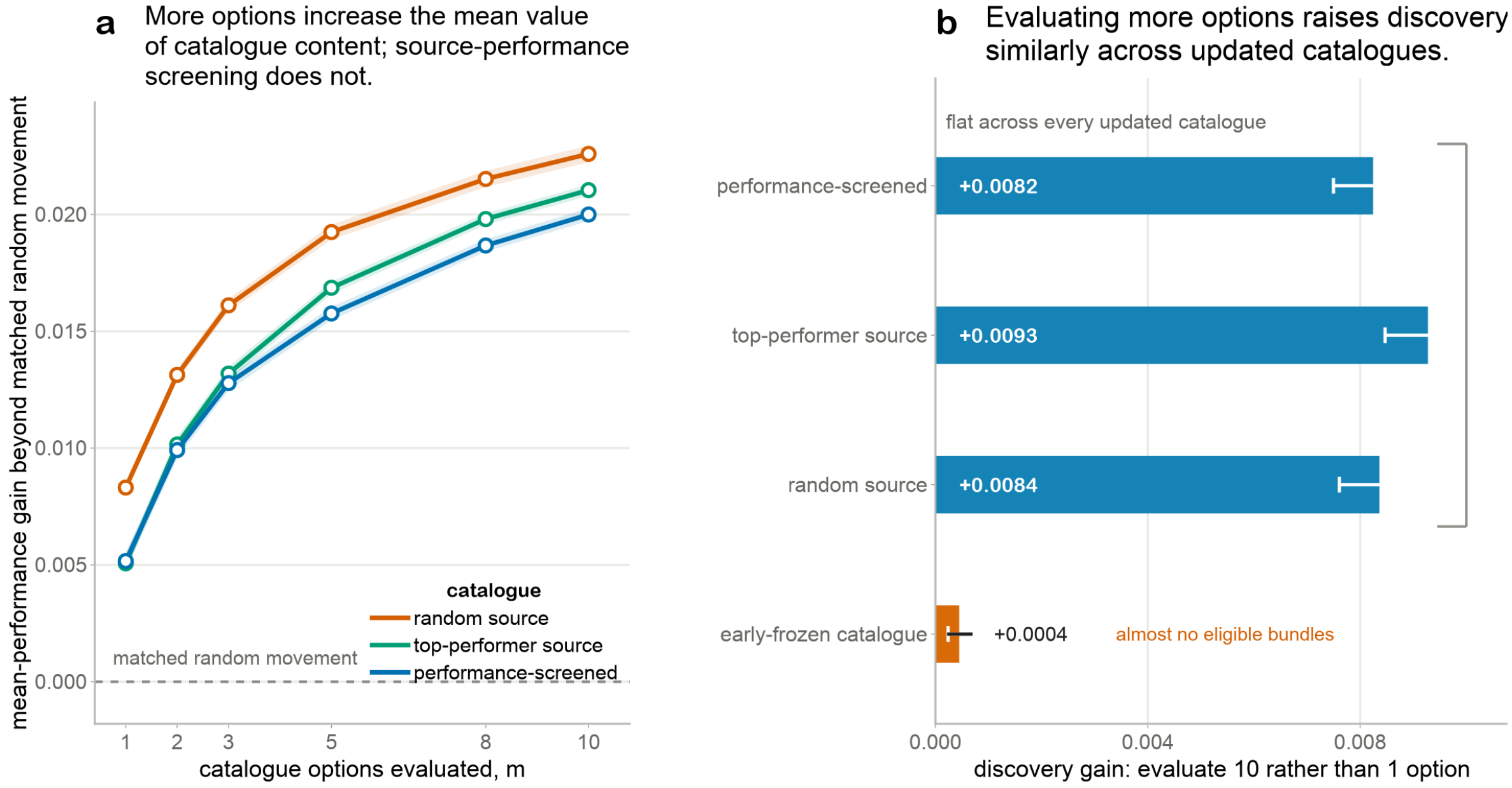

*Figure 3. Catalogue content and the value of comparing alternatives. Top-performer source mines recurring bundles from the ten leading distinct configurations. Performance-screened starts with those bundles but retains only those positively associated with performance among the source configurations. Random source mines recurring bundles from ten randomly selected distinct configurations. The early-frozen benchmark fixes its catalogue during the initial rounds; the list remains, but almost no bundle later satisfies the adopter-specific eligibility rule. (a) Performance advantage of locally selected catalogue content over matched random movement in average firm performance, at six evaluation-capacity levels. Bands are 95% intervals. (b) Change in frontier discovery - the best solution found - when evaluation capacity rises from m = 1 to m = 10. N = 48; static landscape; n = 1,000 paired replications.*

Figure 3a shows a simple pattern: the average-performance advantage of selected catalogue content over matched random movement grows when firms can compare more alternatives. It rises from 0.0051 when a firm sees one catalogue candidate to 0.0210 when it evaluates ten, with diminishing increments. Yet screening bundles by source performance does not produce the largest advantage. At m = 10, the estimate is 0.0226 for random source, 0.0210 for top-performer source, and 0.0200 for performance-screened. Source-performance screening therefore does not increase the total value of selected catalogue content in this comparison.

Figure 3b turns to frontier discovery and asks how the best solution found changes when firms can compare ten catalogue candidates rather than seeing only one. The discovery gain is +0.0082 ± 0.0007 for performance-screened, +0.0093 ± 0.0008 for top-performer source, and +0.0084 ± 0.0008 for random source. The range across the three updated catalogues is only 0.0011, similar to the interval half-widths. In the early-frozen benchmark, the list of old bundles still exists, but almost none remains eligible because its early source configurations no longer outperform current adopters. The m = 10 minus m = 1 return is consequently only +0.0004 ± 0.0002: nominal capacity to evaluate ten candidates does little when the usable choice set is nearly empty.

The comparison separates source selection from publication screening. Top-performer source versus random source changes which configurations supply the raw material. Performance-screened holds the top-performer source fixed and changes which mined bundles are published. Neither rule establishes that a bundle caused its source's performance or guarantees that it will fit a particular adopter; that local fit is what firm judgment evaluates. The random-source catalogue's higher catalogue-content advantage in Figure 3a therefore does not mean practices from random firms are better. It largely reflects more candidates that firms are right to refuse.

The matched random-movement condition provides a final decomposition. Relative to solitary local search, being moved at the same rate and by the same number of decisions adds 0.0048. Locally selected catalogue content adds a further 0.0210 relative to that matched movement. Thus more than four fifths of the focal architecture's total gain comes from which content firms select, not merely from leaving a local peak. Section 5 examines how disruption changes that content advantage.

### 4.2 Population diversity creates the option differences that judgment uses

Where do useful differences among options come from? We vary how similar firms are when search begins and then measure realized population diversity—the share of firms occupying distinct configurations—before any disruption. Figure 4a relates that observed diversity to the additional value of evaluating ten rather than one candidate. We use realized diversity rather than the model's initialization parameter so that the comparison has the same meaning at $N = 32$, 48, and 64.

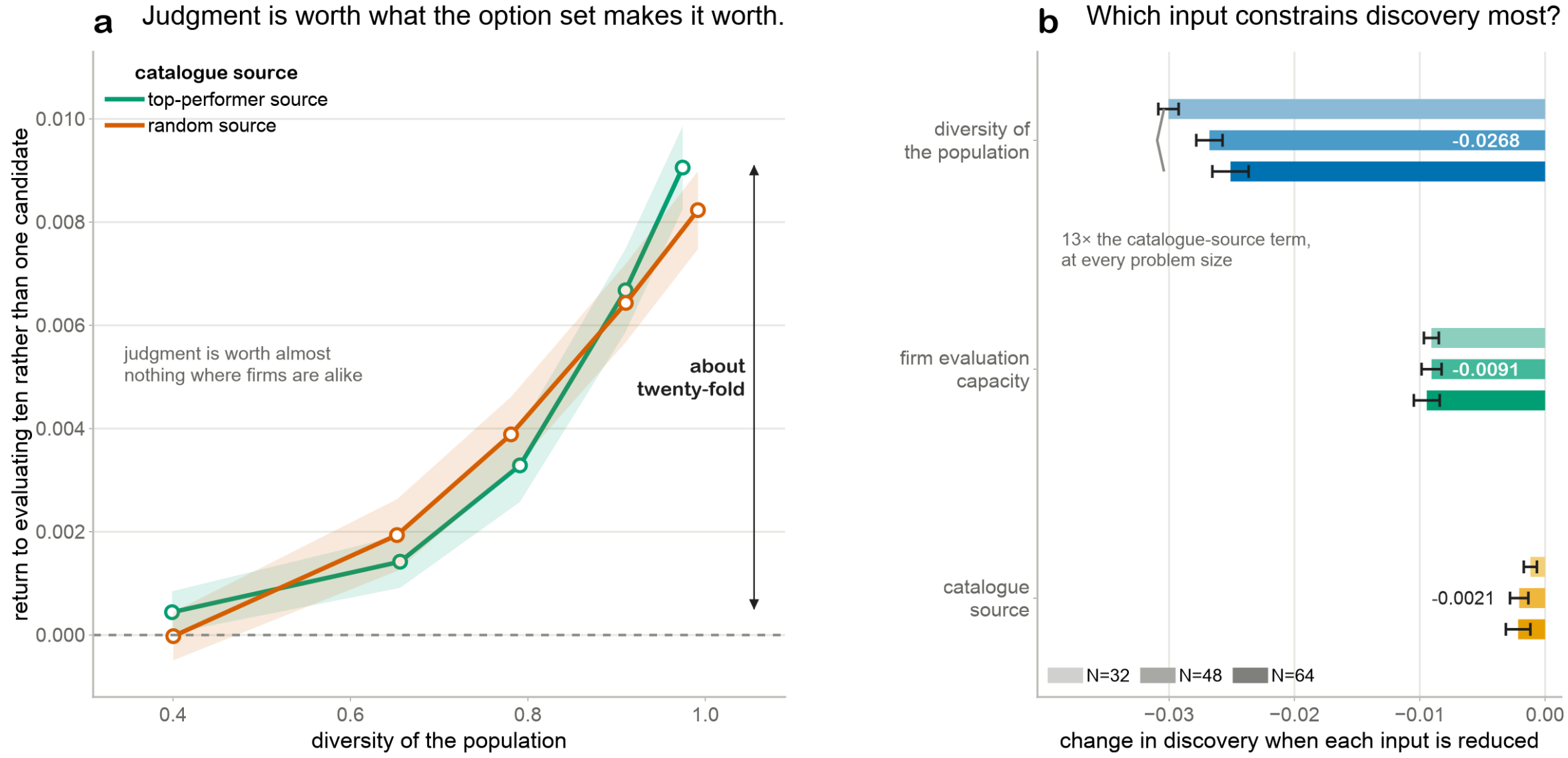

*Figure 4. Population diversity as the source of useful option differences. (a) Additional value of evaluating ten rather than one catalogue candidate, plotted against the realized share of distinct firm configurations for the top-performer and random-source catalogues. Bands are 95% intervals. (b) Change in discovery when population diversity is reduced to its lowest modeled level, evaluation capacity is reduced from ten candidates to one, or catalogue construction is changed from top-performer sources to random sources, at three problem sizes. Labels report the N = 48 estimates.*

The relationship is strong. In the most homogeneous industries, evaluating ten candidates rather than one adds only +0.0004 ± 0.0004. In the most diverse industries, it adds +0.0091 ± 0.0008—about twenty times as much. The same gradient appears with top-performer and random sources. When firms are nearly identical, their shared experience produces candidates that fit them in similar ways, leaving little for judgment to choose among. A diverse population supplies the repertoire with more meaningfully different possibilities.

Figure 4b asks which link in this chain is most constraining. One at a time, the model (1) makes the population as homogeneous as the design permits, (2) reduces evaluation capacity from ten candidates to one, or (3) replaces top-performer sources with random sources. It then measures the resulting change in the best solution discovered. Here, a modeled floor means the lowest level implemented in the experiment; it is not a claim about a natural or real-world minimum.

| input driven to its floor | N = 32 | N = 48 | N = 64 |
|---|---|---|---|
| **diversity of the population** | -0.0300 ± 0.0008 | -0.0268 ± 0.0010 | -0.0251 ± 0.0014 |
| firm evaluation capacity | -0.0091 ± 0.0006 | -0.0091 ± 0.0008 | -0.0094 ± 0.0010 |
| catalogue source | -0.0012 ± 0.0005 | -0.0021 ± 0.0007 | -0.0022 ± 0.0010 |

Every intervention reduces discovery, and the ordering is the same at N = 32, 48, and 64. At N = 48, making the population nearly homogeneous lowers discovery by 0.0268 ± 0.0010. Reducing evaluation capacity lowers it by 0.0091 ± 0.0008, and replacing top-performer sources with random sources lowers it by only 0.0021 ± 0.0007. In this experiment, the population-diversity change is therefore about thirteen times as consequential as the source change and about three times as consequential as the evaluation-capacity change. These ratios describe the modeled interventions; they are not universal estimates.

The conclusion is that population diversity is the scarcest upstream input in this experiment. Selecting sources from top performers or screening bundles by source performance cannot compensate for an industry that contributes too little variation to its shared repertoire.

### 4.3 Judgment creates private value and cannot be replaced by a common rule

Does judgment become a public benefit once many firms possess it? Figure 5a varies the share of firms that can evaluate ten candidates and reports outcomes separately for firms with and without that capacity. Every firm can access the same public catalogue. The comparison asks whether capable firms improve the repertoire enough to benefit firms that cannot evaluate several alternatives themselves.

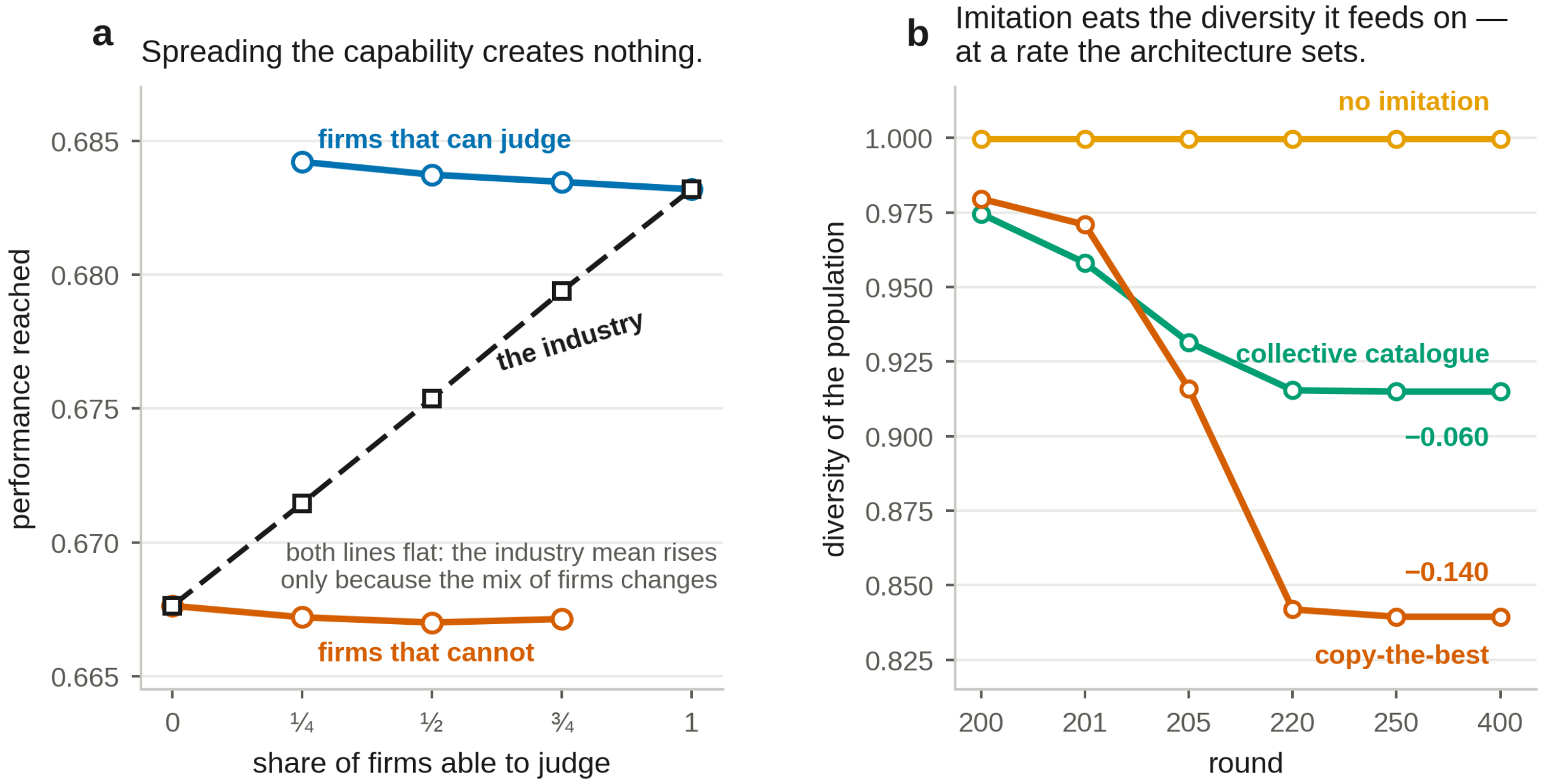

Figure 5. Private judgment and the preservation of population diversity. (a) Performance of firms with and without the capacity to evaluate ten candidates as the capable share of the population increases. (b) Share of distinct firm configurations during the second stage under solitary search, copy-the-best imitation, and learning from a collective catalogue with local evaluation. Conditions share the same horizon, social-learning allocation, and landscapes.

No spillover appears. As the capable share rises from one quarter to three quarters, capable firms remain near 0.684 in mean performance, incapable firms remain near 0.667, and the gap stays close to 0.017. The industry mean rises from 0.6715 to 0.6794 only because more firms belong to the higher-performing group, not because the firms without evaluation capacity improve.

The reason is that judgment concerns a private match: whether a particular practice improves a particular firm's configuration. Making this capability more common changes who benefits from the catalogue, but it does not transfer the capability's return to firms that still lack it.

A common decision rule cannot substitute for that private judgment. We progressively replace local choice with a mandate: firms must adopt the bundle with the greatest catalogue support rather than the candidate that best fits their own configuration. Performance declines whenever the mandate becomes more extensive. When every social adoption is mandated, discovery is 0.0099 ± 0.0008 lower than under free choice, and evaluation capacity has no value because firms are no longer allowed to choose among alternatives.

Surprisingly, the common mandate does not make firms more alike in this model. When every social adoption is mandated, each firm still occupies a distinct configuration (1.000 versus 0.922 under free choice), and the average number of decisions on which two firms differ rises from 18.1 to 22.6. The imposed bundle fits firms differently, so they repair the mismatch in different local directions. The mandate is costly because it destroys the match between firm and practice, not because it reduces configurational diversity.

### 4.4 Learning architectures preserve population diversity at different rates

Today's learning changes the population that will supply tomorrow's repertoire. Figure 5b follows the share of distinct firm configurations through a long period without disruption. It compares three arrangements. Firms either search alone by changing one decision at a time, imitate the best observed peers, or evaluate practices drawn from a collective catalogue built from several groups of leading configurations. All three conditions use the same landscapes and simulation length, and the two social-learning conditions receive the same allowance for imitation.

The arrangements preserve diversity very differently. Solitary search keeps the share of distinct configurations near its ceiling. Under copy-the-best imitation, that share falls from 0.980 to 0.839, a loss of 0.140. Under the collective catalogue with local evaluation, it falls from 0.975 to 0.915, a loss of 0.060—less than half as large.

The mechanisms explain this ordering. Copy-the-best repeatedly pulls firms toward overlapping exemplars. The collective catalogue instead contains practices abstracted from different subsets of leaders, and local evaluation allows firms consulting the same repertoire to choose different candidates. This comparison does not isolate source architecture alone, because the conditions also differ in how bundles are constructed and how adoption is selected. It does show that learning arrangements consume industry heterogeneity at different rates.

The consequences take time to appear. At round 200, the collective-catalogue and copy-the-best conditions do not yet differ detectably. By round 400, the collective-catalogue condition supports both higher discovery and greater population diversity. Its advantage is therefore cumulative, consistent with slower depletion rather than a larger immediate payoff.

This result closes the generativity chain. Population diversity supplies heterogeneous experience; the collective repertoire turns that experience into options; situated judgment extracts value from their different local fit; and the learning architecture determines how quickly adoption consumes the diversity needed to repeat the process. Section 5 uses disruption to reveal the adaptive state that this history of learning has created.

## 5. Results II - What disruption changes and why its effect depends on diversity

We first define the event in landscape terms. Each firm configuration receives a performance score from the settings of its N decisions and the interactions among them. At round 200, disruption redraws the payoff contribution and dependency links for a selected share of those decisions. Firms do not move when this happens: the same configuration is suddenly evaluated under new payoffs and new interdependencies.

The parameter $\varphi$ records how much of the landscape is redrawn. At $\varphi = 0.10$, one tenth of its components change; at $\varphi = 1.00$, all of them change. The expected similarity between pre- and post-disruption performance is $1 - \varphi$. Unless stated otherwise, $N = 48$ and catalogue inertia $\lambda = 0$, meaning that the main results do not deliberately retain old catalogue entries after the event.

This landscape change creates two opposing forces. First, it depreciates prior knowledge: practices that fit before the event may no longer fit afterward. Second, it reopens search: configurations that were local peaks can acquire improving one-step neighbors. The remainder of Section 5 measures both forces and shows why their balance depends on the diversity produced before disruption.

### 5.1 Disruption makes the old repertoire less informative

We begin with knowledge depreciation. Figure 6a tracks the catalogue-content advantage: the average-performance gain from locally selecting among eligible catalogue bundles rather than making a matched random move of the same size at the same rate. Separate lines show five disruption magnitudes, from redrawing 10% of the landscape ($\varphi = 0.10$) to redrawing all of it ($\varphi = 1.00$), plus an undisrupted control. If pre-disruption practices remain informative, this advantage should persist; if the disruption destroys their local fit, it should fall.

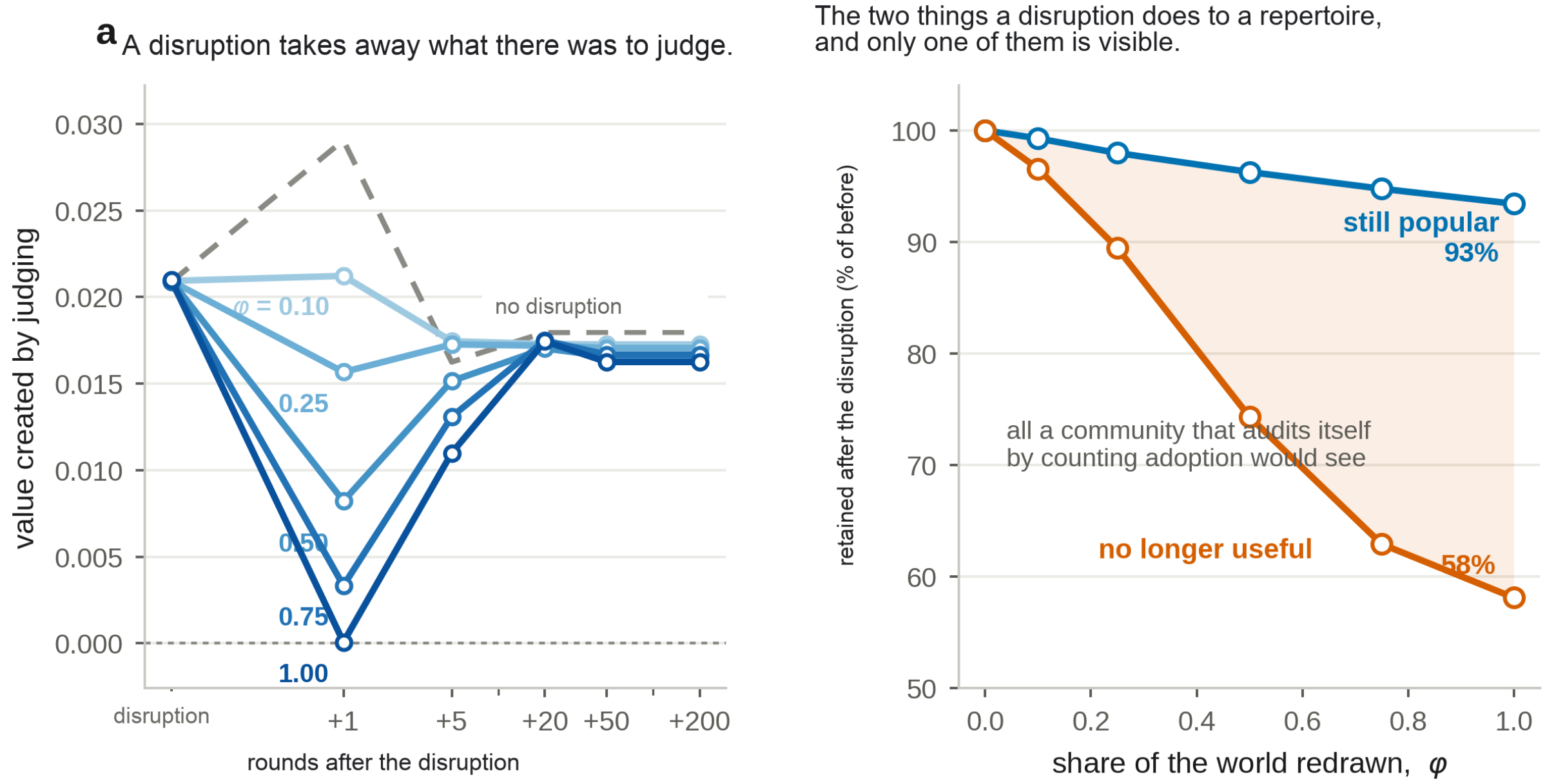

*Figure 6. How disruption weakens a collective repertoire before adoption patterns change. (a) Average-performance advantage of locally selected catalogue content over matched random movement around round 200, by the share of the landscape redrawn; the dashed line is the undisrupted control. A complete redraw removes the immediate value of comparing pre-disruption catalogue options. (b) Immediately after disruption, the share of published practices still recurring among leading configurations and the diversity of those practices' local fit, each indexed to its pre-disruption value.*

The catalogue-content advantage falls immediately, and larger landscape changes produce larger declines. From a pre-disruption value of 0.0210, it is 0.0212, 0.0157, 0.0082, 0.0033, and 0.00007 as the redrawn share rises from 0.10 to 1.00. Under a complete redraw, evaluating ten catalogue candidates is no better in that round than evaluating one. The catalogue still contains recognizable practices, but their pre-disruption content no longer predicts which option fits a firm's unchanged configuration on the new landscape.

The loss is not permanent because firms begin generating post-disruption experience. The catalogue-content advantage becomes positive again within about five rounds and approaches its former level within twenty. It nevertheless remains lower after a complete redraw: 0.0162 versus 0.0180 in the undisrupted control at the same round. The system rebuilds a useful repertoire, but it does so from a population already shaped by earlier convergence and subsequent repair.

Terminal discovery shows the same diminishing pattern. Relative to no disruption, the loss is 0.0030 ± 0.0008 when 10% of the landscape is redrawn and grows to about 0.008 when 75-100% is redrawn. Beyond $\varphi = 0.75$, little additional loss remains because most of the old signal has already been removed. The same decline in option diversity and in the catalogue-content advantage appears at $N = 32$ and $N = 64$.

### 5.2 Popularity survives longer than usefulness

A recurrence-based catalogue observes whether leading configurations still contain a practice; it does not directly retest whether adopting that practice still improves performance. Figure 6b compares these two dimensions immediately after disruption. Support is the share of published practices still recurring among the leaders. Option diversity measures how differently the remaining candidates fit firms locally. Both are expressed relative to their pre-disruption values.

The two indicators separate sharply. After a complete landscape redraw, 93.4% of published practices still satisfy the catalogue's support rule, yet diversity in their local fit falls to 58.1% of its previous level, from 0.0431 to 0.0250. Even at $\varphi = 0.25$, 97.9% of practices retain support. The gap grows monotonically with disruption magnitude. The easiest observable—the fact that firms or leaders still use a practice—is precisely the indicator that reacts least to disruption.

The reason is temporal. A bundle remains published because current leading configurations share its locus settings. Disruption changes the payoff and interdependence attached to those settings immediately, before firms have time to revise their configurations. A once-useful practice can therefore remain common even after the landscape has changed what that practice is worth.

We make this lag consequential by deliberately retaining part of the old catalogue after disruption. On an unchanged landscape, retention raises average performance among firms using local evaluation by $+0.0013 \pm 0.0003$. After a complete redraw, it lowers average performance by $0.0038 \pm 0.0003$, and the cost increases with $\varphi$. This reversal concerns the average firm. It does not appear in the best solution discovered: in the tested comparisons, retention does not switch from improving to reducing the terminal frontier. The evidence therefore supports a change in how performance is distributed across firms, not a reversal of the industry's frontier outcome.

This conclusion is specific to recurrence-based curation. A repertoire that directly retests practices on current performance could detect depreciation; a repertoire maintained mainly because practices continue to appear may not.

### 5.3 Disruption creates new local search opportunities

Knowledge depreciation is only one side of disruption. A firm at a local peak has no one-component change that improves performance on the current landscape. By changing payoffs and interdependencies while leaving the firm's configuration fixed, disruption can turn that former peak into a point with improving neighbors. Figure 7a measures this search reopening through the number of improving moves firms complete with and without a complete landscape redraw.

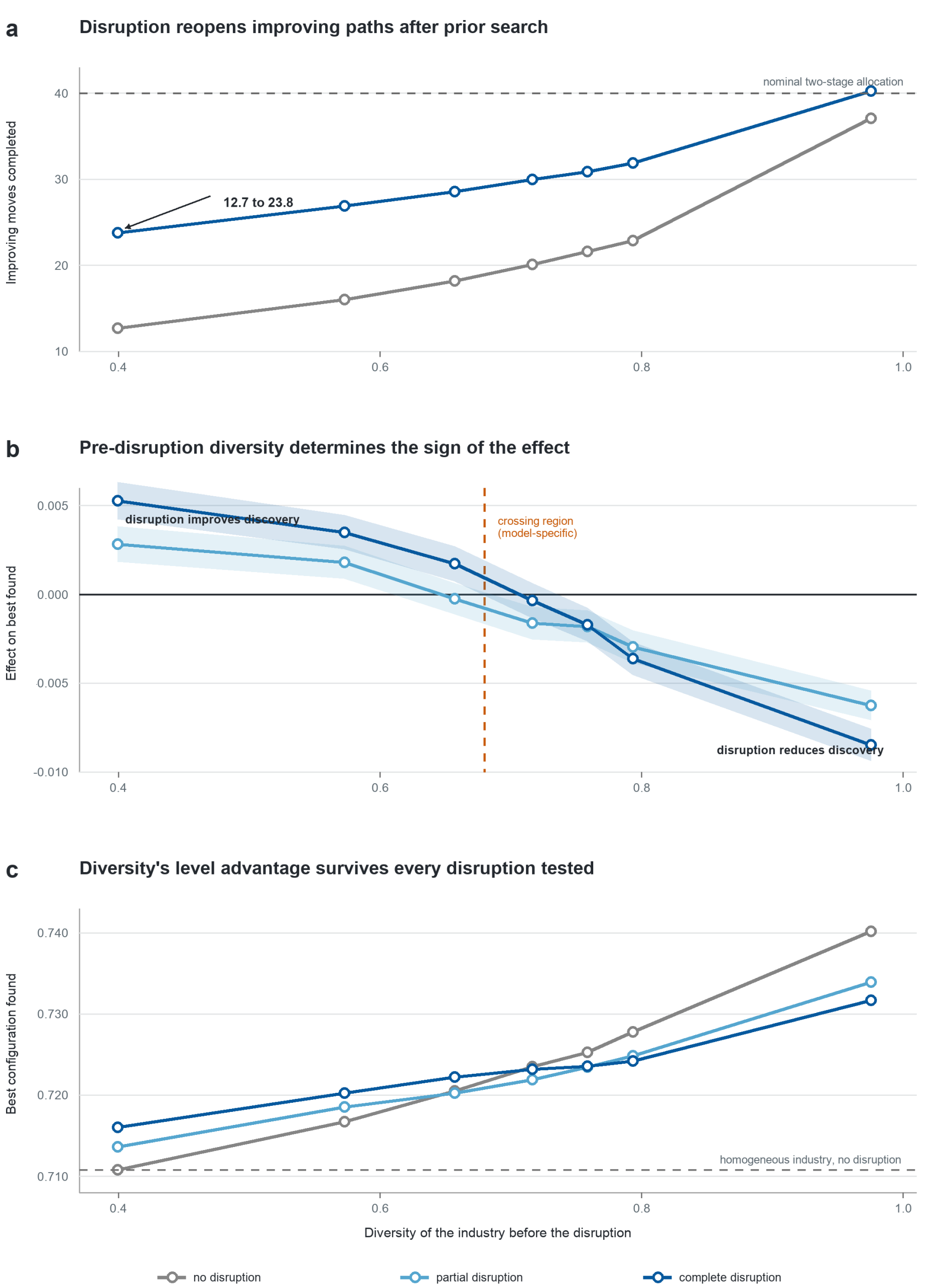

*Figure 7. How disruption reopens search and why its net effect depends on prior diversity. (a) Improving one-decision moves completed with and without a complete landscape redraw; the dashed line marks the nominal allowance of 40 changed decisions across the two stages, not a hard limit on local improvement in the benchmark. (b) Complete disruption minus no*

*disruption in the best solution discovered across seven levels of pre-disruption population diversity; bands are 95% intervals. (c) Final discovery levels for the same conditions.*

Without disruption, low-diversity industries exhaust nearby improvements quickly: firms complete only 12.69 improving moves across the two stages. More diverse populations encounter more reachable improvements, so the count rises to 16.04 and 18.20 as realized diversity increases from 0.40 to 0.66. These are actual improving moves completed, not fractions of a fixed allowance.

A complete landscape redraw nearly doubles movement in the lowest-diversity condition, from 12.69 to 23.78 improving moves. Direct inspection shows why: across the five $K$ values, 98.3% to 100% of firms cease to be local peaks immediately after the redraw; the pooled share is 99.5%. At the most diverse end, the count rises from 37.08 to 40.27. The slight overshoot of 40 confirms that this number is an accounting reference, not a hard ceiling on hill climbing.

Newly available improvements do not automatically produce a better terminal frontier. In a benchmark condition with local search but no locally evaluated repertoire, a complete redraw changes the best solution found by only +0.0005 ± 0.0011 at low diversity and by approximately zero at medium and high diversity. Disruption clearly creates improving neighbors, but one-decision hill climbing alone does not reliably convert those local opportunities into superior final discovery.

We therefore separate two channels. The first is displacement: firms make random moves at the same rate and of the same size as catalogue users, but the moves contain no repertoire information. The second is catalogue content: we compare locally selected catalogue bundles with those matched random moves. This tells us what useful content adds beyond simply leaving a local peak.

The displacement effect at low diversity is not robust; it ranges from -0.0011 to +0.0037 across the four ways of limiting search. The catalogue-content advantage is consistent. After disruption it rises at low diversity in all four designs (+0.0021 to +0.0038) and falls at high diversity in all four (-0.0063 to -0.0161). The stronger evidence therefore concerns how disruption changes the value of collective knowledge, not movement by itself.

An alternative explanation is simple scattering: perhaps disruption helps only because it makes a converged population more diverse. The two explanations make different diagnostic predictions. If scattering drives the result, replications with larger disruption-induced increases in population diversity should show larger discovery gains. If reopened opportunity drives it, replications with larger increases in improving moves should show larger gains.

The scattering prediction fails. Across the three least-diverse conditions and two disruption magnitudes, none of the six correlations between increased population diversity and discovery gain is positive; they range from -0.131 to -0.018, and two are significantly negative. Simply spreading firms apart therefore does not explain the discovery gain.

The reopened-opportunity prediction fares better. All six correlations between increased improving movement and discovery gain are positive, ranging from +0.038 to +0.124, and four are statistically significant. Movement and discovery are partly linked by construction under hill climbing, so these correlations are supporting evidence rather than a stand-alone causal test; the direct loss of local-peak status and the matched comparisons carry more weight.

Together, the diagnostics reject a simple scattering account and support renewed reachability. The landscape redraw removes firms from old local peaks and makes improving one-step neighbors available. Whether firms convert that opportunity into frontier discovery still depends on interdependence, search resources, and repertoire content. These checks support the proposed mechanism, but they do not constitute an experiment that independently manipulates each link.

### 5.4 The net effect of disruption turns negative as prior diversity rises

Figure 7b combines the two forces. For each of seven pre-disruption diversity conditions, it subtracts the best solution discovered without disruption from the best solution discovered after disruption. A positive value means that reopened search more than offsets the loss of prior repertoire value; a negative value means that knowledge depreciation dominates. We report both a partial disruption and a complete landscape redraw.

The effect declines steadily with prior diversity: it is positive in homogeneous, locally exhausted industries and negative in diverse industries with a still-useful repertoire. For a complete disruption, the estimated change crosses zero when about 67.6% of firms occupy distinct configurations; the 95% uncertainty interval runs from 62.6% to 72.0%. This is a summary of the modeled design, not a universal threshold. Other diversity measures produce the same ordering but use different scales; Appendix E reports those estimates.

The primary crossing estimate treats the seven designed diversity conditions as the experimental units and gives more weight to conditions estimated more precisely. This avoids treating random variation among replications within the same condition as if it were a separate diversity treatment. Two alternative calculations place the crossing at 0.6620 and 0.7064; both lie inside the primary interval. Appendix E gives the estimation details.

The sign reversal is not confined to one problem size. Across $N = 32$, 48, and 64, complete disruption raises frontier discovery at the low-diversity endpoint by +0.0033 to +0.0053 and lowers it at the high-diversity endpoint by 0.0085 to 0.0086; all six intervals exclude zero. At $N = 48$, the low-diversity effect is positive at every K under the benchmark and the high-diversity effect is negative at every K. The important exception appears under the hard 20-decision limit: at K = 4 and K = 12, the low-diversity estimates become negative but remain statistically indistinguishable from zero.

We state the crossing in realized diversity because the same initialization setting does not create the same population at different problem sizes. One nominal setting produces realized diversities of 0.2156, 0.3994, and 0.4832 at $N$ = 32, 48, and 64. A threshold in the initialization parameter would therefore be incomparable across sizes. The current design brackets the crossing only coarsely at $N = 32$ and $N = 64$ and does not establish an exactly transferable cutoff.

### 5.5 Disruption changes the marginal effect, not the ranking of industries

The crossing concerns a treatment effect—the difference between disruption and no disruption within the same diversity condition. It does not say that homogeneous industries finish ahead of diverse ones. Figure 7c therefore reports the final discovery levels behind the changes in Figure 7b.

| | homogeneous | middling | diverse | diverse − homogeneous |
|---|---|---|---|---|
| **no disruption** | 0.7108 | 0.7278 | 0.7402 | +0.0294 ± 0.0011 |
| φ = 0.25 | 0.7136 | 0.7249 | 0.7340 | +0.0203 ± 0.0010 |
| φ = 1.00 | 0.7160 | 0.7242 | 0.7317 | +0.0157 ± 0.0010 |

Four comparisons prevent the treatment effect from being mistaken for a performance ranking.

First, greater diversity always raises the final discovery level. All eighteen comparisons between adjacent diversity conditions are positive across no disruption, $\varphi = 0.25$, and $\varphi = 1.00$. Most are statistically significant; under a complete redraw the curve becomes flatter, but it never slopes downward.

Second, a complete redraw narrows but does not eliminate the diversity advantage. The discovery gap between the most and least diverse industries falls from 0.0294 without disruption to 0.0157 after complete disruption, leaving 53% of the original difference.

Third, the diverse industry after a complete redraw still finishes 0.0209 ± 0.0011 above the homogeneous industry with no disruption. It also remains 0.0039 ± 0.0009 above the undisrupted middle-diversity condition.

Fourth, at the homogeneous end, the smallest designed increase in diversity improves discovery by 0.0060 ± 0.0009—about the same as the 0.0053 ± 0.0011 benefit of a complete disruption in that condition. The comparison is descriptive of the model's scale, but it contrasts improving the industry's endogenous state with receiving an exogenous disruption.

The bottom line is therefore not that low diversity is desirable. Disruption can temporarily help a homogeneous industry escape local exhaustion, but diversity raises the frontier throughout. Prior diversity reduces the incremental benefit of disruption while still improving where the industry ultimately finishes.

### 5.6 The conditional effect survives alternative ways of limiting search

The benchmark limits imitation according to the number of decisions changed by an adopted bundle, but it allows an improving one-decision local move to finish even when that move takes the stage total slightly beyond the nominal allowance. We test whether the results depend on this convention in three ways. Two hard-cap variants count every decision changed by local search, social adoption, and subsequent repair, enforcing limits of 20 or 40 changed decisions per stage and rejecting any social bundle that would cross the limit. A fourth design removes the decision-change limit and extends each stage to 400 rounds. Automated checks find no hard-cap violations, and the reimplemented benchmark reproduces the archived outputs exactly. Full audits appear in the online appendix and replication package.

All four search-limit designs reproduce the pooled main contrast: complete disruption raises terminal frontier discovery at low diversity and reduces it at high diversity. Across the four designs, the low-diversity estimates range from +0.0025 to +0.0071, whereas the high-diversity estimates range from -0.0163 to -0.0085. Each estimate averages 1,000 paired replications across five prespecified K values. Figure 8 shows the same decline across all seven diversity conditions. The boundary condition is the hard 20-decision limit: its separate low-diversity effects at $K = 4$ and $K = 12$ are negative but statistically indistinguishable from zero.

The outcome measure also matters. Under every way of limiting search, the high-diversity industry retains a higher terminal frontier than the low-diversity industry, with or without disruption. Yet under the hard limit of 20 changed decisions per stage at low diversity, complete disruption raises the best solution found by +0.0025 while lowering average firm performance by 0.0082. The result concerns the industry's discovered frontier; it does not imply that the average firm benefits from disruption.

The disruption effect remains conditional under four search limits

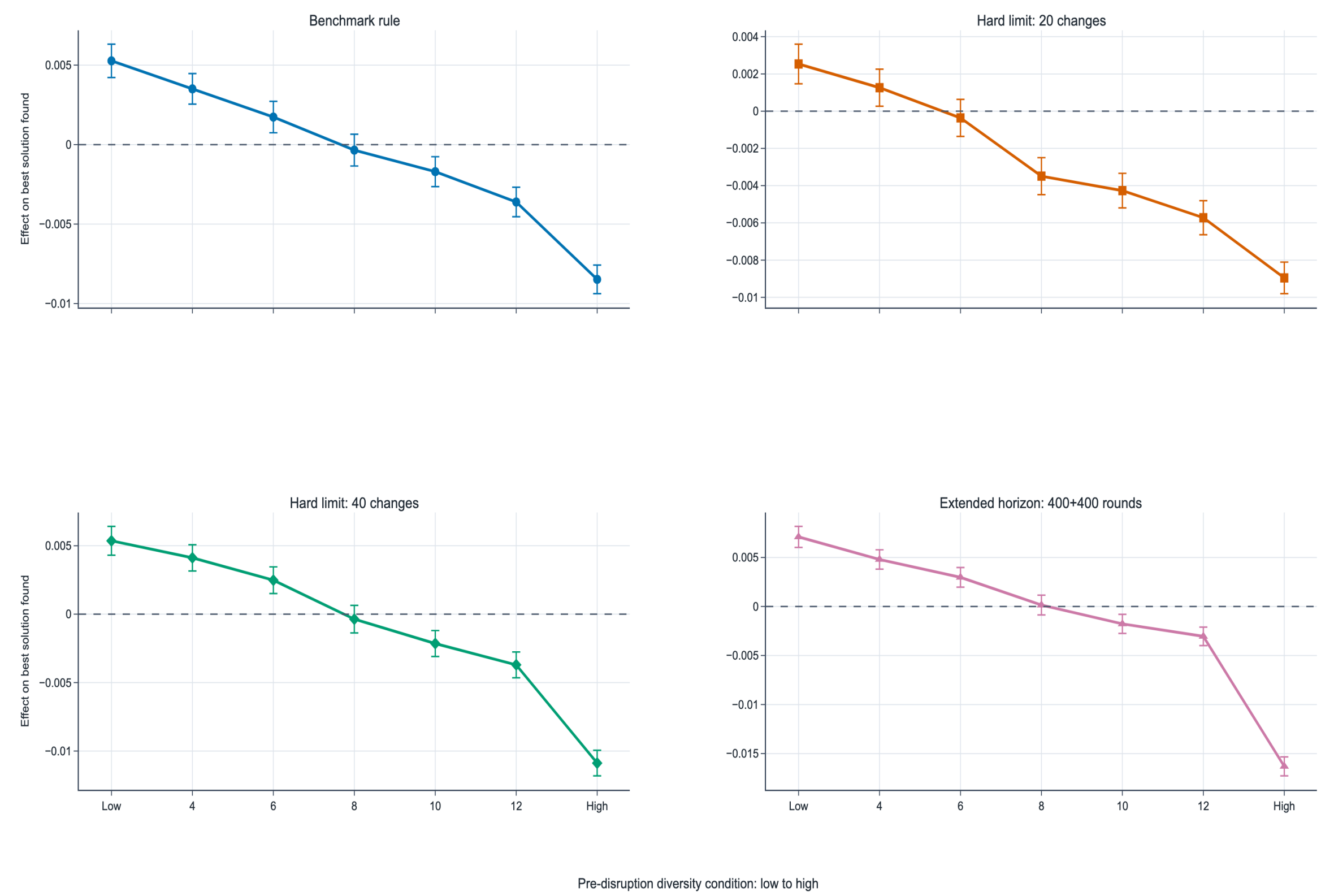

*Figure 8. Robustness of the conditional disruption effect. Complete disruption minus no disruption in terminal frontier discovery across seven designed pre-disruption diversity conditions. Panels compare the archived benchmark, hard limits of 20 and 40 changed decisions per stage, and an extended horizon of 400 rounds per stage without a decision-change limit. Intervals are paired 95% Student-t intervals; dashed lines mark zero.*

## 5.7 Alternative diversity measures and direct mechanism checks

The seven designed conditions order population diversity consistently under three measures. Distinct share is the fraction of firms occupying unique configurations. Normalized pairwise Hamming distance is the average fraction of decisions on which two firms differ. Locus entropy records, decision by decision, how evenly the population is split between the two possible values and then averages that balance. We calculate the same distance and entropy measures among the ten leading source configurations. Because the abstraction rule deliberately selects distinct sources, their distinct share is always one and cannot distinguish the seven conditions; we report that failed measure rather than substituting another after seeing results.

Across these conditions, more population diversity accompanies greater distance among the leading source configurations, more differences in which decisions the published bundles specify, and more dispersion in the performance those bundles would produce for adopting firms. These are descriptive links among outcomes of the same designed upstream condition, not causal mediation estimates. Figure 9 therefore concentrates on the two diagnostics closest to the proposed mechanism: whether disruption creates improving neighbors and whether it changes the value of locally judged catalogue content.

Complete disruption creates improving one-step neighbors at both diversity endpoints. At low diversity, the share of firms with an improving neighbor rises by +0.832 to +0.995 across the four search-limit designs; at high diversity, it rises by +0.360 to +0.995. Under the two hard caps, reopening is stronger at low diversity; under the benchmark and extended-horizon designs, nearly every firm leaves its old peak at both endpoints. These measurements establish that the local gradient changes before firms resume search, but they do not by themselves identify what causes terminal discovery.

The advantage of locally selected repertoire content moves in opposite directions across the two adaptive states. At low diversity, disruption increases the advantage of locally evaluated catalogue content over matched random movement in every search-limit design (+0.0021 to +0.0038). At high diversity, it reduces that advantage in every design (-0.0063 to -0.0161). This is a difference-in-differences: it compares the disruption-induced change in the catalogue condition with the corresponding change in its random-movement benchmark. It is not a full experiment that independently varies old, newly mined, and currently validated content.

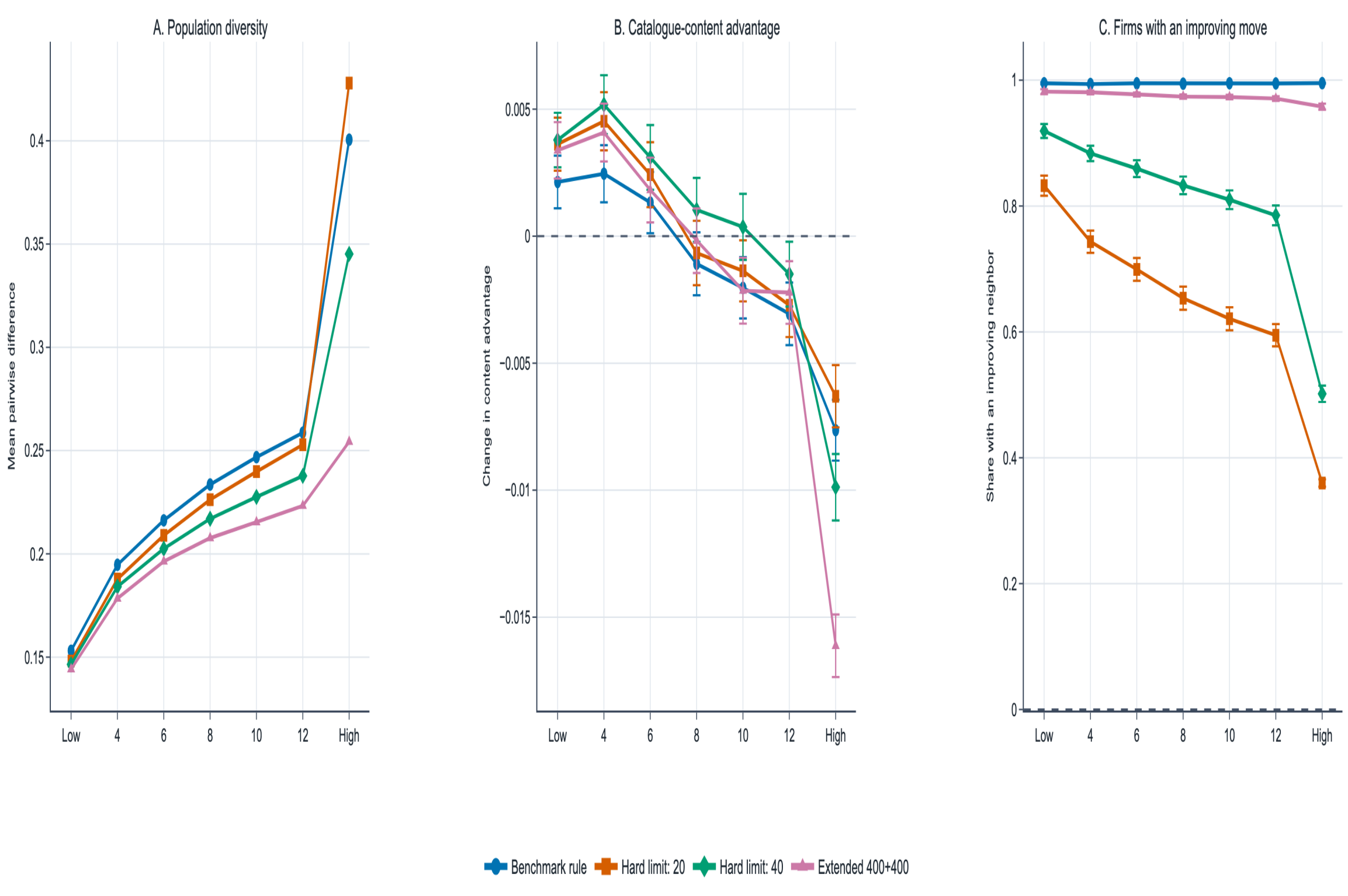

*Figure 9. Direct diagnostics of the conditional disruption mechanism. Panel A reports normalized population Hamming distance before disruption. Panels B and C report how complete disruption changes, respectively, the advantage of locally evaluated catalogue content over matched random movement and the share of firms with an improving one-decision neighbor. Lines connect the seven designed diversity conditions; intervals are paired 95% Student-t intervals where applicable.*

## 6. Discussion

### 6.1 Collective generativity and learning under turbulence

This study explains why the average effect of the same environmental disruption on frontier discovery can be positive in one adaptive state and negative in another. The unifying insight is that the asset and the exposure are the same quantity. Prior diversity creates differences in how catalogue options fit particular firms, which makes situated judgment valuable; a disruption can then depreciate that accumulated usefulness. Prior convergence removes much of that asset and may leave more scope for renewed search, but whether reopened opportunity improves the frontier depends on interdependence, resources, and usable repertoire content.

The principal contribution is a conditional theory of creative destruction in which the effect of an external event depends on an endogenous adaptive state. Population diversity supplies heterogeneous configurations; collective abstraction converts recurring elements among leading configurations into portable best practices; option diversity raises the return to situated judgment; and imitation feeds back on the population. The model thereby shows how prior learning jointly shapes the knowledge exposed to depreciation and the opportunity structure that disruption reopens.

This account extends work that treats diversity primarily as a property of problem solvers. Diversity does broaden the population's search, but its downstream consequence is equally important: it creates consequential alternatives for firms to evaluate. The model accordingly distinguishes diversity in the population from diversity in the local value of the options that a firm sees. The distinction clarifies why improving average catalogue quality cannot substitute for the absence of heterogeneous experience in the field.

The supporting contribution is to theories of organizational search in changing environments. Turbulence does not merely move the target and depreciate knowledge; it also changes which local improvements are reachable. Complete disruption dislodges almost every firm from its prior local peak, but this reopening does not by itself guarantee a better terminal frontier. Renewed reachability is therefore an opportunity whose value is conditional on the surrounding learning system, not exploration by another name.

That conditional effect must be separated from performance levels. In every reported condition, more diverse industries finish ahead. Under a complete disruption, the most diverse industry loses more relative to its own counterfactual yet still outperforms the least diverse industry without a disruption. Creative destruction can be locally restorative without making a depleted adaptive state desirable.

### 6.2 Implications for firms

For firms, the value of evaluation capability depends on the option set supplied by the surrounding field. At low realized population diversity, increasing evaluation capacity from one to ten candidates adds only +0.0004; at high diversity it adds +0.0091. Investment in analytics, experimentation, or managerial judgment should therefore be assessed jointly with the breadth of the alternatives those capabilities can examine.

The model also cautions against delegating contextual decisions to a industry-wide consensus. A widely supported practice can still fit firms differently, particularly when choices are interdependent. Shared access to a repertoire does not eliminate the need for local evaluation, and increasing the prevalence of capable firms does not automatically transfer their gains to firms that lack that capability.

These implications are diagnostic rather than prescriptive estimates. Managers would need empirical measures of relevant differences among firms' configurations, differences in how available practices fit those firms, and exhaustion of nearby improvements before applying the modeled contrasts to a particular industry. The practical question is not whether diversity is high in the abstract, but whether the field continues to supply alternatives with meaningfully different local fit.

### 6.3 Implications for collective knowledge institutions

For the communities and institutions that abstract best practices, curation should preserve variation as well as screen for quality. At N = 48, reducing population diversity to its lowest modeled level lowers discovery by about thirteen times as much as replacing top-performer sources with random sources. A standards body, trade association, or consultancy that publishes only a narrow consensus may improve apparent coherence while reducing the range over which member firms can exercise judgment.

Recurrence is also a weak audit after disruption. Under a complete disruption, 93.4% of practices retain their support even as option diversity falls to 58.1% of its pre-disruption value. Institutions that infer validity from continued adoption can therefore miss rapid depreciation. Retesting consequences in current configurations is more informative than asking only whether a practice remains common.

The Basel Committee's post-2007–08 restrictions on banks' internal models, including a 72.5% aggregate output floor, and manufacturers' efforts to reassess just-in-time supply systems after the pandemic illustrate the governance problem (Basel Committee on Banking Supervision 2017; Zhang and Doan 2021). They are not tests of this model. They show why organizations that codify recurring practice may need mechanisms for renewed local validation after disruption.

Collective knowledge institutions face a boundary they cannot solve through filtering alone: no curation rule can recover configurational experience that a homogeneous field no longer contains. Maintaining multiple sources, retaining minority practices long enough to evaluate them, and recording conditions of use are ways to keep that limitation visible.

### 6.4 Implications for policy and institutional design

For policy makers, the central lever is the architecture of learning rather than disruption itself. Copying the best observed firm reduces population diversity by 0.140 over the second stage, compared with 0.060 when firms select locally from a collectively sourced catalogue. Institutions that encourage comparison among several sources while preserving local discretion may sustain the variation on which future adaptation depends.

The results do not justify inducing disruptions or protecting homogeneity. In the model, the smallest designed increase in diversity at the homogeneous end produces a gain comparable to the benefit of a complete disruption, while the diverse industry facing the complete disruption still outperforms the undisrupted homogeneous industry by +0.0209. Building a generative state is the controllable objective; disruption is not.

Generative artificial intelligence offers a contemporary illustration. A common model may improve individual output while drawing users toward overlapping solutions (Doshi and Hauser 2024). The model here suggests evaluating such systems not only by immediate average quality but also by whether their sourcing and use preserve heterogeneous options for later judgment. This is an analogy, not an empirical claim about a particular technology.

Across these applications, the policy principle is the same: measure and govern the production of options, not merely the diffusion of favored practices. The relevant outcomes include the variety of the contributing population, dispersion in local fit, and the continued availability of improving moves.

## 7. Limitations and boundary conditions

The model isolates adaptive discovery rather than appropriated profit, survival, or incumbent displacement. Firms do not enter, exit, set prices, or respond strategically to rivals, and demand does not migrate. The findings therefore concern how a population searches, learns, and rebuilds shared knowledge after a

disruption—not every process associated with technological disruption. Empirical applications should distinguish changes in discovery from changes in market position or value capture.

The crossing is a conditional model result, not a universal diversity threshold. Its location depends on the landscape, search rule, population, learning architecture, and diversity measure. What travels is the mechanism: prior diversity shapes both the option value exposed to depreciation and the local opportunities a disruption reopens. Appendix F reports the robustness checks; Appendix G records the measurement and implementation boundaries behind them.

The mechanism comparisons sharpen interpretation but do not fully identify every link. Matched random movement separates useful repertoire content from displacement, yet the current design does not independently cross retained pre-disruption content, newly mined content, currently validated content, and the presence of situated judgment. The popularity–usefulness gap is likewise specific to repertoires curated through recurrence. Institutions that continuously retest current consequences, deliberately retain minority experience, or attach context to practices may be less vulnerable.

The model also holds firms' response rules and learning architecture fixed. Real organizations may reconfigure networks, change how practices are produced, invest in new evaluation capabilities, or strategically retain old practices. Empirical work can test the argument by tracing whether pre-disruption differences among firms generate dispersion in how shared practices fit them, whether disruption changes that dispersion before prevalence changes, and whether renewed local search converts into discovery. These observations would distinguish the generativity mechanism from resilience, scattering, and simple knowledge depreciation.

## 8. Conclusion

Collective learning is usually judged by the improvements it spreads today. That standard is incomplete. A learning system also changes the population from which tomorrow's practices will be generated. This paper shows that the same architecture that distributes improvement can preserve or consume the differences required for future discovery.

First, we develop a theory of collective generativity. Population diversity supplies heterogeneous experience; collective abstraction converts that experience into a repertoire of reusable options; situated judgment matches those options to local configurations; and subsequent imitation changes the raw material available for the next cycle. This chain reframes diversity as a productive input to knowledge creation, not only a buffer against shocks. It also explains why access to shared knowledge cannot substitute for local evaluation: a common repertoire becomes valuable only when firms can judge fit.

Second, the analysis identifies learning architecture as a source of long-run adaptive capacity. Copying leading firms improves performance quickly but narrows the population that future abstraction can draw on. A collectively sourced repertoire combined with local judgment preserves more variation and therefore more generative capacity. The contribution is not that collective learning must homogenize. It is that architectures differ in whether today's learning replenishes or depletes tomorrow's option supply.

Third, the paper offers a conditional account of disruption and creative destruction. Disruption both depreciates accumulated collective knowledge and reopens local improvement paths. The same event can therefore aid a homogeneous, locally exhausted industry while harming a diverse industry with more useful knowledge to lose. Yet the diverse industry can still finish ahead. This effect-versus-level distinction changes how diversity should be interpreted: greater exposure to loss is not evidence that diversity is disadvantageous; it may be the consequence of having built a more valuable adaptive asset.

Fourth, the findings expose a governance problem for firms and policy makers. Popularity can outlive usefulness after disruption, so continued adoption is a poor validity test. Firms should pair shared repertoires—including AI-enabled ones—with current, local tests of fit and should monitor whether the breadth of options is shrinking. Policy makers and professional institutions should preserve heterogeneous sources, retain contextual information, and evaluate learning systems by the options they continue to produce, not only by the practices they diffuse. The objective is neither to induce disruption nor to reject common knowledge. It is to organize collective learning so that improving today does not imply narrowing tomorrow.

## References

Afuah, A., C. L. Tucci. 2012. Crowdsourcing as a solution to distant search. Academy of Management Review 37(3) 355–375.

Almirall, E., R. Casadesus-Masanell. 2010. Open versus closed innovation: A model of discovery and divergence. *Academy of Management Review* 35(1) 27–47.

Argote, L., S. Lee, J. Park. 2021. Organizational learning processes and outcomes: Major findings and future research directions. Management Science 67(9) 5399–5429.

Basel Committee on Banking Supervision. 2017. *Basel III: Finalising post-crisis reforms*. Bank for International Settlements, Basel, 7 December.

Cennamo, C., J. Santaló. 2019. Generativity tension and value creation in platform ecosystems. Organization Science 30(3) 617–641.

Clement, J., P. Puranam. 2018. Searching for structure: Formal organization design as a guide to network evolution. Management Science 64(8) 3879–3895.

Competition and Markets Authority. 2024. AI Foundation Models: Update Paper. UK Competition and Markets Authority, London, 11 April.

Csaszar, F. A., N. Siggelkow. 2010. How much to copy? Determinants of effective imitation breadth. *Organization Science* 21(3) 661–676.

Denrell, J. 2003. Vicarious learning, undersampling of failure, and the myths of management. *Organization Science* 14(3) 227–243.

Denrell, J., J. Luukkonen, N. Chater, C. Liu. 2026. The need for absorptive capacity alleviates the free-rider problem in knowledge production. Management Science, Articles in Advance. https://doi.org/10.1287/mnsc.2024.06471.

Doshi, A. R., O. P. Hauser. 2024. Generative AI enhances individual creativity but reduces the collective diversity of novel content. *Science Advances* 10(28) eadn5290.

Ethiraj, S. K., D. Levinthal. 2004. Modularity and innovation in complex systems. *Management Science* 50(2) 159–173.

Fang, C., J. Lee, M. A. Schilling. 2010. Balancing exploration and exploitation through structural design: The isolation of subgroups and organizational learning. *Organization Science* 21(3) 625–642.

Gavetti, G., D. Levinthal. 2000. Looking forward and looking backward: Cognitive and experiential search. *Administrative Science Quarterly* 45(1) 113–137.

Henderson, R. M., K. B. Clark. 1990. Architectural innovation: The reconfiguration of existing product technologies and the failure of established firms. *Administrative Science Quarterly* 35(1) 9–30.

Hong, L., S. E. Page. 2004. Groups of diverse problem solvers can outperform groups of high-ability problem solvers. *Proceedings of the National Academy of Sciences* 101(46) 16385–16389.

Jensen, R. J., G. Szulanski. 2007. Template use and the effectiveness of knowledge transfer. Management Science 53(11) 1716–1730.

Kauffman, S. A. 1993. *The Origins of Order: Self-Organization and Selection in Evolution*. Oxford University Press, New York.

Knudsen, T., D. A. Levinthal. 2007. Two faces of search: Alternative generation and alternative evaluation. *Organization Science* 18(1) 39–54.

Lazer, D., A. Friedman. 2007. The network structure of exploration and exploitation. *Administrative Science Quarterly* 52(4) 667–694.

Lee, D., E. Van den Steen. 2010. Managing know-how. Management Science 56(2) 270–285.

Levinthal, D. A. 1997. Adaptation on rugged landscapes. *Management Science* 43(7) 934–950.

March, J. G. 1991. Exploration and exploitation in organizational learning. *Organization Science* 2(1) 71–87.

Page, S. E. 2007. *The Difference: How the Power of Diversity Creates Better Groups, Firms, Schools, and Societies*. Princeton University Press, Princeton, NJ.

Posen, H. E., D. A. Levinthal. 2012. Chasing a moving target: Exploitation and exploration in dynamic environments. *Management Science* 58(3) 587–601.

Posen, H. E., J. Lee, S. Yi. 2013. The power of imperfect imitation. *Strategic Management Journal* 34(2) 149–164.

Rivkin, J. W. 2000. Imitation of complex strategies. *Management Science* 46(6) 824–844.

Schumpeter, J. A. 1942. *Capitalism, Socialism and Democracy*. Harper & Brothers, New York.

Siggelkow, N., J. W. Rivkin. 2005. Speed and search: Designing organizations for turbulence and complexity. *Organization Science* 16(2) 101–122.

Villarroel, J. A., J. E. Taylor, C. L. Tucci. 2013. Innovation and learning performance implications of free revealing and knowledge brokering in competing communities: Insights from the Netflix Prize challenge. Computational and Mathematical Organization Theory 19(1) 42–77.

Weitzman, M. L. 1998. Recombinant growth. *Quarterly Journal of Economics* 113(2) 331–360.

Winter, S. G., G. Szulanski. 2001. Replication as strategy. Organization Science 12(6) 730–743.

Zhang, H., T. T. H. Doan. 2021. From just-in-time to just-in-case: Global sourcing and firm inventory after the pandemic. *VoxEU/CEPR column*, Centre for Economic Policy Research, London.

# Online Appendix for "Improving Today, Narrowing Tomorrow: Collective Learning, Diversity, and Generativity"

## A. Formal specification

### A.1 Landscape

An industry of M = 100 firms searches a common NK landscape. A configuration x is a binary string of length N. Each locus i has a dependency set D_i containing K other loci, drawn uniformly at random and fixed for the run, and a contribution function c_i defined over its own value and the values of its K dependencies.

For each of the 2^(K+1) input combinations, c_i is drawn independently from Uniform(0,1).

The performance of configuration x is f(x) = (1/N) Σ_i c_i(x_i, x_{D_i}).

We report *N* = 48 in the main text and replicate at $N \in \{32,64\}$. *K* runs over a five-point grid spanning nearly separable to nearly fully coupled landscapes. Note the convention: *N* counts loci; the firm count is *M* and is always 100.

### A.2 The round

Write $x_t^{(j)}$ for firm j's configuration at round $t$ and $\mathcal{N}_1(x)$ for its single-locus neighborhood. A round proceeds:

1. **Local improvement.** If there exists $y \in \mathcal{N}_1\left(x_t^{(j)}\right)$ with $f(y) > f\left(x_t^{(j)}\right)$, the firm moves to the best such y, charging one move.
2. **Stuck.** Otherwise the firm is at a local peak, and social imitation becomes available.
3. **Source.** The firm draws a candidate from its assigned source (§A.3).
4. **Evaluation.** A firm with capacity *m* draws *m* candidates and evaluates $f(\cdot)$ at the configuration each would produce, adopting the best if it improves on the status quo. A firm with *m* = 1 evaluates one; a blind firm adopts without evaluating.
5. **Repair.** After adoption the firm resumes local improvement from wherever the adoption left it.

Every changed locus is recorded in the move counter. Social imitation is available while a firm's stage allocation remains; improving hill-climbing moves are completed and recorded even if they take the count slightly above that allocation. We set the social-learning allocation to 20 per stage with one renewal at t = 200. All arms in a comparison face the same horizon and allocation, and realized movement is reported explicitly.

Priority robustness protocol. The current policy reproduces the archived gate exactly. Under the strict-locus policies, a local-search or repair flip costs one locus, a social adoption costs its realized Hamming displacement, and every realized configuration change is admitted only if the full cost fits within the stage allocation; multi-locus moves are atomic and never partially applied. Allocations are 20 or 40 loci and renew at the disruption boundary. The uncapped policy records the same actual-locus ledger without an affordability gate and extends both stages from 200 to 400 rounds. All variants use N=48, M=100, K in {4,12,24,36,46}, 200 replications per K, the same seed architecture and disruption draws, and paired no-, partial-, and complete-disruption cells.

### A.3 Sources

Let $\mathcal{P}_j$ be firm j's observation neighborhood (ten firms, fixed). Write $\mathcal{C}_t$ for the published catalog at round *t* (§B).

| arm | source | adoption |
|---|---|---|
| `M1` | none | — (solitary search) |
| `M2` | the best-performing peer in the neighborhood | blind: copy a random subset of that peer's loci |
| `M2_EVAL_SUBSET(m)` | as `M2` | evaluate m subsets of that peer, adopt the best |
| `M2_EVAL_PEER(m)` | m peers from the neighborhood | evaluate one subset from each, adopt the best |
| `A_EVAL(1)` | the catalog at round t | evaluate one drawn bundle, adopt if it improves |
| `A_EVAL(m)` | the catalog at round t | evaluate m drawn bundles, adopt the best if it improves |
| `A_MATCH(m)` | the catalog at round t | **matched placebo**: adopt a bundle at random, matched to `A_EVAL(m)` on firing rate and bundle size |

`A_MATCH(m)` is the comparator for every content claim in the paper. It fires as often as `A_EVAL(m)`, takes bundles of the same size from the same catalog, and is matched on spend before any outcome is read. The difference between `A_EVAL(m)` and `A_MATCH(m)` is therefore the value of *judging*, net of the value of *being moved*.

### A.4 Outcomes and interpretation

The primary outcome is discovery, measured as the best configuration found. Population-mean performance and population diversity are secondary outcomes reported separately. A mean-only gain is not described as a discovery gain, particularly when it coincides with a lower best-found value or a collapse in diversity.

### A.5 Seeding and statistics

The landscape, the initial population, the observation network and the disruption are seeded from the tuple (run seed, $N$, $K$, replication index) only. Arm-level stochasticity adds the arm and condition index. The same landscapes therefore appear in every condition, and contrasts are paired exactly on $(K,\text{replication})$.

For a paired contrast with $n$ pairs, we report $\bar{d} \pm t_{0.975,n-1}\, s_d/\sqrt{n}$. Degrees of freedom are always $n-1$; we never use a normal approximation.

## B. The abstraction operator and what it can publish

### B.1 Definition

Let E_t = {e_1,...,e_q}, with q at most E = 10, denote the highest-performing distinct configurations present at round t after duplicate firm configurations are collapsed. A pattern is a partial assignment P = {(i,v_i): i in I}, where I is a subset of the N loci and v_i is binary. Configuration e exhibits P if e_i = v_i for every i in I. The support of P is

$$\text{supp}(P) \;=\; |\{\, j \in \mathcal{E}_t \,:\, j \text{ exhibits } P \,\}|.$$

A pattern is *closed* if no strict superset of it has the same support. The catalog is the set of closed patterns with $\text{supp}(P) \geq s_{\min}$ and $|I| \leq L_{\max}$, where τ = 0.25 gives a support floor of $s_{\min} = 2$ of the 10 leading distinct configurations and $L_{\max} = \lceil N/3 \rceil$; at most $A$ = 100 are retained for firms to draw from. **A bundle therefore requires only a small minority of the elite to share it**, which is why the catalog holds many overlapping entries rather than a single consensus.

Crucially, **no value is published**. The catalog is a set of partial assignments with no fitness attached. A firm learns what a bundle is worth only by evaluating $f(\cdot)$ at the configuration adopting it would produce.

### B.2 The bound

For $S \subseteq \mathcal{E}_t$ non-empty, define the *agreement pattern*

$$A(S) \;=\; \left\{ (i,v) \,:\, x_i^{(j)} = v \;\text{ for every } j \in S \right\}.$$

**Proposition B.1.** *Every closed pattern of $\mathcal{E}_t$ equals* A*(S) for some non-empty $S \subseteq \mathcal{E}_t$. Consequently the number of distinct closed patterns is at most $2^E - 1$, independent of* N.

*Proof.* Let P be closed and let $S_P = \{j \in \mathcal{E}_t : j \text{ exhibits } P\}$, so $\text{supp}(P) = |S_P|$. Every configuration in $S_P$ agrees with P on I, hence $P \subseteq A(S_P)$. And every configuration in $S_P$ exhibits $A(S_P)$ by construction, so $\text{supp}\big(A(S_P)\big) \geq |S_P| = \text{supp}(P)$; since $A(S_P) \supseteq P$ and P is closed, $A(S_P) = P$. The map $S \mapsto A(S)$ therefore surjects the non-empty subsets of $\mathcal{E}_t$ onto the closed patterns, of which there are at most $2^E - 1$.
□

With $E$ = 10 the ceiling is 1,023 patterns however large $N$ is. The mined catalog is thus a bounded object whose size is governed by the elite's internal structure, not by the size of the problem.

### B.3 The raw-material constraint

**Corollary B.2. If all E elite configurations hold the same configuration x, every nonempty subset of elites yields the same agreement pattern. The catalog therefore contains exactly one pattern and offers nothing to choose among.**

This is the formal statement of §2.2. What a community can publish is bounded by what its leading members do *not* have in common. In a homogeneous industry the intersection lattice collapses, the catalog degenerates to a single item, and the firm's capacity to weigh *m* candidates has nothing to weigh. It is why the return to that capacity in our most homogeneous condition is +0.0004 and in our most diverse is +0.0091.

The corollary also explains why a *quality filter* cannot repair a thin catalog. Filtering selects a subset of $A(\cdot)$; it cannot create patterns that the elite's agreement structure does not contain.

## C. Why option diversity and not option quality

### C.1 The general statement

Let $X_1, \ldots, X_m$ be exchangeable random variables representing the fit of *m* candidate bundles to a given firm's configuration, and define the value of the capacity to weigh *m* candidates as

$$V(m) \;=\; \mathbb{E}\left[\max_{1 \leq i \leq m} X_i\right] \;-\; \mathbb{E}[X_1].$$

**Proposition C.1 (location invariance, scale homogeneity). Let Y_i = a + bX_i with b > 0. Then V_Y(m) = bV_X(m) for all m. Thus V is invariant to a pure shift in candidate quality and homogeneous of degree one in scale.**

*Proof. Because b > 0, max_i(a + bX_i) = a + b max_i(X_i). Taking expectations and subtracting the value of one draw gives the result.* ■

The proposition supplies a conditional benchmark. If a catalog manipulation shifts every candidate by a common amount while leaving dispersion and dependence unchanged, it does not change the value of choosing among candidates. Manipulations that also change dispersion, dependence, or the accept-or-reject margin need not satisfy this invariance.

### C.2 The Gaussian case, and a testable shape

If $X_i \sim \mathcal{N}(\mu, \sigma^2)$ independently, then $\mathbb{E}[\max_i X_i] = \mu + \sigma a_m$ where $a_m = \mathbb{E}[\max_{i \le m} Z_i]$ for standard normal $Z_i$, so

$$V(m) \;=\; \sigma\, a_m, \qquad a_m = m \int_{-\infty}^{\infty} z\ \phi(z)\, \Phi(z)^{m-1}\, dz.$$

Under the independent Gaussian benchmark, the constants imply that the ladder in m is linear in the expected-normal-maximum term a_m, with slope σ. The benchmark therefore offers a descriptive shape against which to compare the simulated ladder; it is not a distributional assumption of the model.

### C.3 The fit

Regressing the measured content premium at $N = 48$ on $a_m$ across the six values of $m$:

| catalog | intercept | slope ($\hat{\sigma}$) | $R^2$ |
|---|---|---|---|
| performance-screened | 0.00480 | 0.00966 | 0.9969 |
| top-performer source | 0.00464 | 0.01053 | 0.9970 |
| random source | 0.00816 | 0.00939 | 0.9989 |

Three things are worth drawing out.

**The Gaussian benchmark describes the observed shape closely: the regression $R^2$ exceeds 0.997 in each catalog condition. This high fit is useful as a compact summary of the ladder, but it does not validate Gaussian, independence, or location-scale assumptions outside these simulated cells.**

**The fitted slope is similar across catalog conditions—0.0094 to 0.0105, a range of 0.0011. Within this benchmark, the slope summarizes the scale on which candidate bundles differ in fit. The result is consistent with, rather than a direct measurement of, Proposition C.1's location-scale mechanism.**

**The intercept captures the single accept-or-reject margin available at m = 1. It is 70% larger for the random-source catalog (0.0082 versus 0.0046–0.0048), consistent with that catalog containing more**

**candidates that a firm is right to reject. This decomposition is descriptive: the random-source catalog does not appear to offer a wider option set, but it offers more to refuse.**

**Why the population's diversity governs σ.** By Corollary B.2, the candidate bundles are agreement patterns of the elite. As the elite becomes more alike, those patterns converge on a single item and the differences in fit across drawn candidates shrink; σ → 0 and, by C.2, $V(m) \to 0$ for every $m$. The measured ladder falls from +0.0091 to +0.0004 across our diversity range, which is what that limit predicts.

## D. The disruption operator

### D.1 Definition

At round $T_d = 200$ a set $D \subseteq \{1, \ldots, N\}$ with $|D| = \lfloor \varphi N \rceil$ is drawn uniformly. For each $i \in D$:

- **weights variant:** the contribution table $c_i$ is re-drawn independently from the same distribution;
- **weights-and-structure variant (reported throughout):** the dependency set $B_i$ is re-drawn uniformly *and* the table $c_i$ is re-drawn on the new arguments.

No firm's configuration is altered; what changes is what configurations are worth. This is a shift in the payoff environment of the kind Posen and Levinthal (2012) study, with the architectural component that Henderson and Clark (1990) emphasize added in the second variant.

### D.2 The correlation between the pre- and post-disruption landscape

**Proposition D.1.** *Let f and f' denote performance before and after the disruption. For a configuration x drawn independently of the tables,*

$$\mathrm{Corr}\big(f(x),\, f'(x)\big) \;=\; 1 - \varphi.$$

*Proof. The pre- and post-disruption performance functions share the contribution terms for loci outside D; contributions for loci in D are independent redraws. With independent locus contributions and common variance σ_c², the two functions share exactly N(1 − φ) terms, so*

$$\mathrm{Cov}(f, f') = \frac{1}{N^2}\sum_{i \notin D} \mathrm{Var}\,(c_i) = \frac{(1-\varphi)\,\sigma_c^2}{N}, \qquad \mathrm{Var}(f) = \mathrm{Var}(f') = \frac{\sigma_c^2}{N},$$

and the covariance-to-variance ratio is 1 − φ. ■

We derived this before reading any output from the disruption module and it is confirmed to four decimal places at every φ we run, in both variants. It is the one piece of the implementation validated against an analytic result rather than against itself, and we report it as such.

### D.3 Catalog inertia

The parameter $\lambda \in [0,1]$ is the fraction of the published catalog carried over unrevised after the disruption, retaining its *recorded* support — that is, the support it had before the world changed. Setting λ = 0 re-mines the catalog from scratch each round, which is the assumption we use everywhere except §5.2, so that inertia is never confounded with the disruption. Setting λ > 0 models the realistic case in which a repertoire is revised more slowly than the world moves.

## E. Estimation of the crossing

### E.1 The estimand

Let $d_c$ be the realized pre-disruption population diversity of cell c and $y_c$ the paired effect of the disruption on discovery in that cell, with standard error $s_c$. Fit

$$y_c \;=\; \alpha + \beta\, d_c + \varepsilon_c, \qquad w_c = s_c^{-2},$$

by weighted least squares across the seven cells. The crossing is the ratio

$$d^* \;=\; -\hat{\alpha}/\hat{\beta}.$$

### E.2 Why the cell means and not the replications

A per-replication regression has 7,000 observations behind it and looks more authoritative. It is attenuated. Within a cell, realized diversity varies mostly as sampling noise rather than as the designed contrast, so regressing on it is a classical errors-in-variables problem and shrinks $\hat{\beta}$ toward zero. The seven cell means are the designed contrast; each is measured to about 0.008 in diversity and 0.0005 in effect, and weighting by precision uses the design as it was built.

| estimator | φ = 1.00 | φ = 0.25 |
|---|---|---|
| **cell-mean WLS (primary)** | **0.6759** | **0.6265** |
| per-replication OLS | 0.6620 | 0.5803 |
| bracketing interpolation | 0.7064 | 0.6470 |

The three span 0.662 to 0.706 at φ = 1.00, narrower than the primary interval itself, so the choice of estimator does not drive the conclusion. The reportable statement is "about 0.68".

### E.3 Fieller intervals

A ratio of estimated coefficients has no symmetric confidence interval, and a delta-method band understates it when $\hat{\beta}$ is not overwhelmingly precise. We invert the test instead: the 1 - γ confidence set for $d^*$ is

$$\left\{ d : \left(\hat{\alpha} + \hat{\beta} d\right)^2 \ \le\ t^2_{1-\gamma/2,\nu} \left(\hat{V}_{\alpha\alpha} + 2d\hat{V}_{\alpha\beta} + d^2 \hat{V}_{\beta\beta}\right) \right\},$$

a quadratic inequality in *d* whose root interval is reported. At φ = 1.00 this gives [0.6263, 0.7202]; at φ = 0.25, [0.5793, 0.6649].

### E.4 Monotonicity and uniqueness

The response is monotone across all seven levels at both magnitudes, with zero significant sign changes: the effect passes through zero once and does not return. A non-monotone response or a second crossing would complicate the two-process account in Section 2.4; neither appears in these designed cells.

## F. Supporting tables

### F.1 The seven-cell design in levels

*N = 48, weights-and-structure disruption, λ = 0, top-performer-source catalog, arm A_EVAL(10), 400 rounds, two nominal social-learning allocations, n = 1,000 pairs per cell. Diversity is the share of distinct configurations in the φ = 0 control at round 200.*

| pre-disruption diversity | φ = 0 | φ = 0.25 | φ = 1.00 |
|---|---|---|---|
| 0.3994 | 0.71078 | 0.71361 | 0.71605 |
| 0.5729 | 0.71674 | 0.71855 | 0.72025 |
| 0.6570 | 0.72052 | 0.72028 | 0.72225 |
| 0.7163 | 0.72354 | 0.72192 | 0.72319 |
| 0.7583 | 0.72529 | 0.72348 | 0.72358 |
| 0.7934 | 0.72782 | 0.72488 | 0.72421 |
| 0.9752 | 0.74020 | 0.73396 | 0.73173 |

Paired contrasts, most diverse minus most homogeneous: +0.02942 ± 0.00109 at $\varphi = 0$; +0.02035 ± 0.00101 at $\varphi = 0.25$; +0.01568 ± 0.00096 at $\varphi = 1.00$. Diverse-and-fully-disrupted minus homogeneous-and-never-disrupted: +0.02095 ± 0.00109.

### F.2 Improving moves completed and the nominal allocation

| pre-disruption diversity | control | φ = 0.25 | φ = 1.00 |
|---|---|---|---|
| 0.3994 | 12.69 | 18.73 | 23.78 |
| 0.5729 | 16.04 | 22.19 | 26.88 |
| 0.6570 | 18.20 | 24.03 | 28.56 |
| 0.7163 | 20.12 | 25.74 | 29.96 |
| 0.7583 | 21.61 | 27.06 | 30.87 |
| 0.7934 | 22.85 | 27.90 | 31.89 |
| 0.9752 | 37.08 | 38.25 | 40.27 |

### F.3 The two channels, correlated against the gain in discovery

Within-landscape correlations, averaged across the five landscape settings; the three least diverse conditions.

| condition | scattering channel | reopened-opportunity diagnostic |
|---|---|---|
| r = 2, φ = 0.25 | −0.050 | +0.110 |
| r = 2, φ = 1.00 | −0.053 | +0.106 |
| r = 4, φ = 0.25 | −0.131 | +0.124 |
| r = 4, φ = 1.00 | −0.074 | +0.060 |
| r = 6, φ = 0.25 | −0.114 | +0.108 |
| r = 6, φ = 1.00 | −0.018 | +0.038 |

Zero of six positive for scattering; six of six positive for the gradient, significant in four.

### F.4 The flip at three problem sizes

| N | homogeneous: diversity, effect | diverse: diversity, effect | n |
|---|---|---|---|
| 32 | 0.2156, +0.00328 ± 0.00091 | 0.7803, -0.00864 ± 0.00076 | 2500 |
| 48 | 0.3994, +0.00527 ± 0.00105 | 0.9752, -0.00848 ± 0.00090 | 1000 |
| 64 | 0.4832, +0.00474 ± 0.00130 | 0.9999, -0.00848 ± 0.00117 | 500 |

All six significant. Note the realized diversity at the same nominal setting: 0.2156, 0.3994, 0.4832 — a range of 0.27, which is why the threshold is stated in diversity.

### F.5 What a disruption does to the catalog

| φ | practices retaining support | catalog diversity at T+1 | as % of pre-disruption |
|---|---|---|---|
| 0 | 99.9% | 0.04309 | 100% |
| 0.10 | 99.2% | 0.04161 | 96.6% |
| 0.25 | 97.9% | 0.03855 | 89.5% |
| 0.50 | 96.2% | 0.03203 | 74.3% |
| 0.75 | 94.7% | 0.02712 | 62.9% |
| 1.00 | 93.4% | 0.02505 | 58.1% |

### F.6 The cost of a community norm

Forcing intensity h; N = 48, top-performer-source catalog, static landscape, n = 1000.

| h | discovery vs free choice | distinct configurations | mean pairwise distance | return to judging |
|---|---|---|---|---|
| 0 | — | 0.922 | 18.14 | +0.00952 ± 0.00081 |
| 0.25 | -0.00180 ± 0.00070 | 0.938 | 18.38 | +0.00777 ± 0.00076 |
| 0.50 | -0.00306 ± 0.00074 | 0.955 | 18.83 | +0.00673 ± 0.00076 |
| 1.00 | -0.00988 ± 0.00078 | 1.000 | 22.63 | 0.00000 |

### F.7 Robustness of the disruption implementation

The two disruption variants — re-drawing payoffs alone, and re-drawing payoffs while rewiring dependencies — are indistinguishable at every magnitude on every measure we report. The main text uses the weights-and-structure variant throughout. As one instance, the share of practices losing support at φ = 1.00 is 0.0665 under weights and 0.0664 under weights-and-structure; catalog diversity at $T + 1$ is 0.02527 and 0.02505.

The collapse in catalog diversity and in the return to judgment replicates at both other problem sizes: catalog diversity at φ = 1.00 falls to 0.0271 at $N = 32$ and 0.0224 at $N = 64$, with the return to judging falling to zero in both.

### F.8 Registered predictions

Predictions were written into the run drivers before execution and never revised; the driver headers are preserved in the replication package. Where a registered prediction ran against a claim held in the literature, the result is reported in the main text as a result. Registered predictions concerning internal conjectures are not reported as findings; two are reported because they discriminate between competing mechanisms and the discrimination is load-bearing: the scattering channel (§5.3), which the registered prediction expected to hold and which the data contradict, and the direction of the crossing in φ (§5.4).

## G. Additional boundary conditions and implementation details

### G.1 Diversity measures and the crossing

The simulated crossing is expressed in model units and should not be exported as a universal threshold. Distinct share depends on the fixed population of M = 100 firms and approaches its maximum as nearly every firm becomes unique. Normalized pairwise Hamming distance and locus entropy avoid that ceiling, but measure different properties on different scales and consequently yield different fitted crossings. We report each measure separately and estimate a crossing only when the observed conditions lie on both sides of zero. Empirical work should trace diversity among leading sources and dispersion in candidate fit without treating any simulated crossing as universal.

### G.2 Limits of the mechanism comparison

Matched random-movement conditions separate locally selected catalogue content from displacement, but they do not complete causal mediation. Across search-limit designs, the advantage of locally selected content rises at low diversity and falls at high diversity. Displacement alone is positive at low diversity in three designs but not under the hard 20-decision cap. A stronger factorial experiment would independently vary four repertoire conditions—none, retained pre-disruption content, content re-mined from current leaders, and content explicitly validated after disruption—and cross each with the presence or absence of situated judgment. The main text therefore treats the present comparisons as mechanism evidence rather than complete identification.

### G.3 Search budgets and observation horizons

The archived implementation limits social imitation but allows an improving one-decision local move to pass the nominal allowance. Robustness analyses impose hard 20- and 40-decision stage limits on every realized change and also test a 400+400-round horizon without a decision-change limit. Even under the longer horizon, a material share of runs is still improving when observation stops, so we do not describe the endpoint as convergence. These checks address budget headroom and simulation length within the same search architecture; they do not establish robustness to every landscape, network, or behavioral rule.

### G.4 Landscape, population, and network scope

The model uses binary NK landscapes, M = 100 firms, one local-search rule, and a fixed observation network. Results replicate at N = 32, 48, and 64 and across five levels of interdependence. Other landscape structures, population sizes, dynamic networks, and search rules may shift the crossing or remove it. These are scope conditions on the quantitative result, not changes to the proposed two-process logic.

### G.5 Abstraction rule and source diversity

The model compresses distributed abstraction into a transparent recurrence rule applied to up to ten leading distinct configurations. Because those sources are selected to be distinct, the share that is distinct is always one and cannot discriminate among conditions; only their distance from one another and their decision-by-decision diversity are informative. Institutions that test causal performance, deliberately represent minority experience, or preserve contextual information may be less vulnerable to the popularity–usefulness gap. The gap is therefore a boundary condition of recurrence-based curation, not a claim about every knowledge institution.

### G.6 Analytic and replication safeguards

The relationship between disruption magnitude and similarity of the pre- and post-disruption landscapes was derived analytically and then confirmed computationally. Focal mechanism predictions were recorded before the simulations were interpreted, and the complete code and replication materials accompany the paper. These safeguards support implementation validity but do not remove the model's substantive boundary conditions.